\documentclass[aps,prx,floatfix,showpacs,twocolumn, preprintnumbers]{revtex4-1}

\usepackage{caption}
\usepackage{subcaption}
\usepackage{graphicx}
\usepackage{amssymb}
\usepackage{amsmath}
\usepackage{mathtools}
\usepackage[dvipsnames,usenames]{xcolor}
\usepackage{multirow}

\newcommand{\be}{\begin{equation}}
\newcommand{\ee}{\end{equation}}
\newcommand{\bs}[1]{\ensuremath{\boldsymbol{#1}}}

\begin{document}

\preprint{LA-UR-21-32144}

\title{Neutrinoless double-beta decay: combining quantum Monte Carlo and the nuclear shell model with the generalized contact formalism}
\author{ {Ronen} Weiss$^{\, {\rm a} }$,
{Pablo} Soriano $^{\, {\rm b} }$,
{Alessandro} Lovato $^{\, {\rm c,d,e} }$,
{Javier} Menendez $^{\, {\rm b} }$,
{R. B.} Wiringa $^{\, {\rm c} }$,
}
\affiliation{
$^{\,{\rm a}}$\mbox{Theoretical Division, Los Alamos National Laboratory, Los Alamos, New Mexico 87545, USA}\\
$^{\,{\rm b}}$\mbox{Department of Quantum Physics and Astrophysics and Institute of Cosmos Sciences, University of Barcelona, 08028 Barcelona, Spain}\\
$^{\,{\rm c}}$\mbox{Physics Division, Argonne National Laboratory, Argonne, Illinois 60439, USA}\\
$^{\,{\rm d}}$\mbox{Computational Science Division Division, Argonne National Laboratory, Argonne, Illinois 60439, USA}\\
$^{\,{\rm e}}$\mbox{INFN-TIFPA Trento Institute of Fundamental Physics and Applications, Via Sommarive, 14, 38123 {Trento}, Italy}\\
}
\date{\today}

\begin{abstract}
We devise a framework based on the generalized contact formalism that combines the nuclear shell model and quantum Monte Carlo methods and compute the neutrinoless double-beta decay of experimentally relevant nuclei, including $^{76}$Ge, $^{130}$Te, and $^{136}$Xe. In light nuclei, we validate our nuclear matrix element calculations by comparing against accurate variational Monte Carlo results. Due to additional correlations captured by quantum Monte Carlo and introduced within the generalized contact formalism, in heavier systems, we obtain long-range nuclear matrix elements that are about 30\% smaller than previous shell-model results. We also evaluate the recently recognized short-range nuclear matrix element estimating its coupling by the charge-independence-breaking term of the Argonne $v_{18}$ potential used in the Monte Carlo calculations. Our results indicate an enhancement of the total nuclear matrix element by around 30\%.
\end{abstract}

\maketitle

\section{Introduction}
 Neutrinoless double-beta decay ($0\nu\beta\beta$) is a hypothetical transition of atomic nuclei, forbidden by the Standard Model of particle physics, in which two neutrons are transmuted into two protons and two electrons are emitted with no accompanying antineutrinos~\cite{Furry:1939qr}. The measurement of $0\nu\beta\beta$ decay would have profound implications, demonstrating that lepton number is not a symmetry of nature, proving that the neutrino mass has a Majorana component~\cite{Schechter:1981bd}---so that neutrinos and antineutrinos are the same particle---and, since two matter particles are created without the corresponding antimatter ones, illuminating the matter dominance in the universe. In addition, assuming the decay is mediated by the exchange of light neutrinos, it would provide invaluable insight on the neutrino mass scale and ordering~\cite{DellOro:2016tmg,Cardani:2018lje}. 

Because $0\nu\beta\beta$ is a second order decay, in practice it can only be detected in the few nuclei where $\beta$ decay is either energetically forbidden or strongly suppressed by spin change. Current best limits are given for $^{76}$Ge ($T^{0\nu}_{1/2} > 1.8 \times 10^{26}$ yr~\cite{GERDA:2020xhi}) and $^{136}$Xe ($T^{0\nu}_{1/2} > 1.07 \times 10^{26}$ yr~\cite{KamLAND-Zen:2016pfg}), and in this decade next-generation ton-scale experiments plan to reach $T^{0\nu}_{1/2} \sim 10^{28}$ yr, mainly in $^{76}$Ge, $^{100}$Mo, $^{130}$Te or $^{136}$Xe~\cite{nEXO:2018ylp,NEXT:2020amj,LEGEND:2021bnm,CUPID:2019imh}. 
Since the decay involves physics beyond the Standard Model (BSM), its rate is proportional to a parameter describing the lepton number violation in that BSM mechanism. In addition, the decay rate is also governed by a nuclear matrix element (NME) that encodes the structure of the initial and final nuclei. Thus, extracting specific BSM information from half-life measurements demands reliable NMEs. This involves high-quality nuclear structure studies of heavy nuclei such as $^{76}$Ge or $^{136}$Xe. In this work we focus on the light Majorana neutrino exchange, but NMEs for other decay mechanisms can be calculated in the same fashion.

Most NME calculations use the quasiparticle random-phase approximation (QRPA)~\cite{Mustonen:2013zu,Simkovic:2013qiy,Hyvarinen:2015bda,Fang:2015zha}, nuclear shell model (SM)~\cite{Menendez:2008jp,Iwata:2016cxn,Horoi:2015tkc,Coraggio:2020hwx}, energy-density functional theory~\cite{Rodriguez:2010mn,LopezVaquero:2013yji,Yao:2014uta,Yao:2016oxk} or the interacting-boson model~\cite{Barea:2015kwa,Deppisch:2020ztt}. Besides other aspects---for an extensive review see Ref.~\cite{Engel:2016xgb}---these many-body methods show deficiencies related to the inconsistent treatment of the $0\nu\beta\beta$-decay operator. These may appear as missing nuclear correlations or two-nucleon currents in the decay operator. {\it Ab initio} many-body methods, in contrast, treat transition operators consistently, as they describe nuclear properties emerging from the bare interaction between protons and neutrons. This way they reproduce well $\beta$-decay rates in light- and medium-mass nuclei~\cite{Gysbers:2019uyb,King:2020wmp} without any adjustments---usually known as ``quenching''---, a feature required by less sophisticated many-body approaches like those mentioned above~\cite{Chou:1993zz,Wildenthal:1983zz,Martinez-Pinedo:1996zvt,Pirinen:2015sma,Barea:2015kwa}. {\it Ab initio} methods relying on single-particle basis expansion have calculated $0\nu\beta\beta$-decay NMEs for $^{48}$Ca~\cite{Novario:2020dmr,Yao2020,Wirth:2021pij}, or even heavier $^{76}$Ge and $^{82}$Se~\cite{Belley:2020ejd}. On the other hand, quantum Monte Carlo (QMC) approaches are very accurate, but so far limited to light nuclei with up to $A\leq12$ nucleons~\cite{Pastore:2017ofx,Cirigliano:2018hja}. QMC treats accurately short-range nuclear dynamics, a key aspect for $0\nu\beta\beta$ transitions where decaying neutrons are typically a few fm apart. This is even more critical for the recently recognized short-range NME~\cite{Cirigliano:2018hja,Cirigliano:2019vdj}.

The generalized contact formalism (GCF) is a powerful tool to model the short-range behavior of nuclear distributions, both in coordinate and momentum space~\cite{Weiss:2014gua, Weiss:2015mba,Weiss:2016obx,Cruz-Torres2020}. Nuclear wave functions show a universal short-range behavior determined by the nuclear interaction; the only dependence on the specific nucleus comes about as an overall factor, proportional to the number of short-range correlated (SRC) pairs. Hence, the GCF is applicable across the nuclear chart, provided that these normalization factors are known. In this work, we use state-of-the-art variational Monte Carlo (VMC) and SM $0\nu\beta\beta$ transition densities to determine the ratios of normalization factors---which do not depend on the nuclear interaction---and correct the short-range part of the SM transition densities, assuming they describe well the long-range behavior. This way we include the short-range dynamics captured by the VMC into the SM. Although similar in spirit to correcting shell-model transition densities with Jastrow correlations~\cite{Miller_Spencer_1976,Simkovic2009,Benhar:2014cka,Wang:2019hjy}, our approach is more systematic, as it allows us to exactly match VMC results in $A\leq 12$ systems, and can accommodate a variety of nuclear interactions. 

After validating our method in light nuclei where VMC calculations are available, we make NME predictions for nuclei used in $0\nu\beta\beta$ experiments, including $^{48}$Ca, $^{76}$Ge, $^{130}$Te, and $^{136}$Xe. Our results include estimates of the theoretical uncertainty associated with our method.
In addition to the usual long-range NME, we compute the leading-order short-range NME, for which our approach may be particularly reliable. We follow recent analyses and estimate the coupling associated with this term by the charge-independence-breaking (CIB) term of the Argonne $v_{18}$ (AV18) potential used in the VMC calculations.
While for this work we use the phenomenological AV18 plus the three-nucleon Urbana X (UX) force, our method is general and can readily be applied to interactions derived within chiral effective field theory.

The manuscript is organized as follows. First, we introduce the $0\nu\beta\beta$ transition potentials in Sec.~\ref{sec:trans_pot}. Section~\ref{sec:many_body} describes the many-body methods, while Sec.~\ref{sec:results} presents our NME results. Finally, Sec.~\ref{sec:conclusions} summarizes our main conclusions and future perspectives.

\section{$0\nu\beta\beta$ Transition potentials}
\label{sec:trans_pot}
Under the closure approximation~\cite{Pantis:1991va,Faessler:1991qi,Senkov:2014vyd}, the $0\nu\beta\beta$ NME between the initial and final nuclear states $|\Psi_i\rangle$ and $|\Psi_f\rangle$ reads
\begin{equation}
M^{0\nu} = \langle \Psi_f| O^{0\nu} | \Psi_i\rangle\,.
\end{equation}
We focus on the light Majorana neutrino exchange. For this mechanism the long-range transition operator can be cast as a sum of
Fermi (F), Gamow-Teller (GT) and tensor (T) contributions $O^{0\nu}_L= O^{0\nu}_F + O^{0\nu}_{GT} + O^{0\nu}_{T}$, where
\begin{align}
O^{0\nu}_F &= (4\pi R_A) \sum_{a \neq b} V^{0\nu}_F(r_{ab}) \tau_a^+  \tau_b^+\,,\nonumber\\
O^{0\nu}_{GT} &= (4\pi R_A) \sum_{a \neq b} V^{0\nu}_{GT}(r_{ab})\, \sigma_{ab}  \, \tau_a^+  \tau_b^+\,,\nonumber\\
O^{0\nu}_{T} &= (4\pi R_A) \sum_{a \neq b} V^{0\nu}_{T}(r_{ab}) S_{ab} \, \tau_a^+  \tau_b^+\,  .
\end{align}
Here $\bs{\sigma}_a$ and $\tau_a$ represent nucleon spin and isospin operators, respectively, $\sigma_{ab} = \bs{\sigma}_a \cdot \bs{\sigma}_b$, and the tensor operator is $S_{ab}=3(\bs{\sigma}_a\cdot\hat{r}_{ab})(\bs{\sigma}_b\cdot\hat{r}_{ab})-\sigma_{ab}$ with $r_{ab}$ the internucleon distance. The nuclear radius $R_A = 1.2\, A^{1/3}$~fm is inserted by convention to make the NME dimensionless. The coordinate-space neutrino potentials above are obtained from the standard Fourier transform:
\begin{equation}
V_\alpha^{0\nu}(r_{ab}) = \frac{1}{g_A^2} \int \frac{d^3 \mathbf{q}}{(2\pi)^3} e^{i \mathbf{q}\cdot\mathbf{r}_{ab}} V_\alpha^{0\nu}(\mathbf{q}^2)\,,
\end{equation}
where $\mathbf{q}$ is the momentum transfer, $\alpha$ indicates F, GT, and T, and we take $g_A= 1.27$ for the axial-vector coupling.

Defining $V_\alpha^{0\nu}(\mathbf{q}^2) = \frac{1}{\mathbf{q}^2} v_\alpha(\mathbf{q}^2)$ the relevant functions can be given in terms of the nucleon isovector vector, axial, induced pseudoscalar and magnetic form factors~\cite{Pastore:2017ofx,Engel:2016xgb}:
\begin{align}
v_F(\mathbf{q}^2) &= - g_V^2(\mathbf{q}^2)\,, \nonumber\\
v_{GT}(\mathbf{q}^2) &= g_A^2(\mathbf{q}^2) + \frac{2}{3} \frac{\mathbf{q}^2}{2 m_N}g_A(\mathbf{q}^2)g_P(\mathbf{q}^2) \nonumber\\
& + \frac{1}{3} \frac{\mathbf{q}^4}{4 m_N^2} g_P^2(\mathbf{q}^2) + \frac{2}{3} \frac{\mathbf{q}^2}{4 m_N^2} g_M^2(\mathbf{q}^2)\,, \nonumber\\
v_T(\mathbf{q}^2) &= - \frac{2}{3} \frac{\mathbf{q}^2}{2 m_N} g_A(\mathbf{q}^2)g_P(\mathbf{q}^2) - \frac{1}{3} \frac{\mathbf{q}^4}{4 m_N^2}g_P^2(\mathbf{q}^2)\nonumber\\
& + \frac{1}{3}\frac{\mathbf{q}^2}{4 m_N^2} g_M^2(\mathbf{q}^2)\,,
\label{eq:transition_pot}
\end{align}
where $m_N=938.9$~MeV is the nucleon mass. Consistent with the $0\nu\beta\beta$ literature, for the single-nucleon form factors we adopt the simple dipole parameterization:
\begin{align}
g_V(\mathbf{q}^2) &= \frac{g_V}{(1+\mathbf{q}^2/\Lambda_V^2)^2}\,,\nonumber\\
g_M(\mathbf{q}^2) & = (1+\kappa_1)g_V(\mathbf{q}^2)\,,\nonumber\\
g_A(\mathbf{q}^2) &= \frac{g_A}{(1+\mathbf{q}^2/\Lambda_A^2)^2}\,,\nonumber \\
g_P(\mathbf{q}^2) &= -\frac{2 m_N }{q^2+m_\pi^2}g_A(\mathbf{q}^2)\,,
\end{align}
with vector coupling $g_V=1$, anomalous nucleon isovector magnetic moment $\kappa_1=3.7$, and pion mass $m_\pi=138$~MeV. The cutoff values are $\Lambda_V=0.85$ GeV and $\Lambda_A=1.04$ GeV. More sophisticated functional forms for these form factors exist~\cite{Ye:2017gyb}, including some based on a systematic z-expansion~\cite{Meyer:2016oeg}. However, for the relatively small momentum transfer at play in $0\nu\beta\beta$ processes, $|\mathbf{q}|\sim200$~MeV, no significant differences are expected with respect to the simple dipole ansatz.

\begin{figure}[t]
\includegraphics[width=\linewidth]{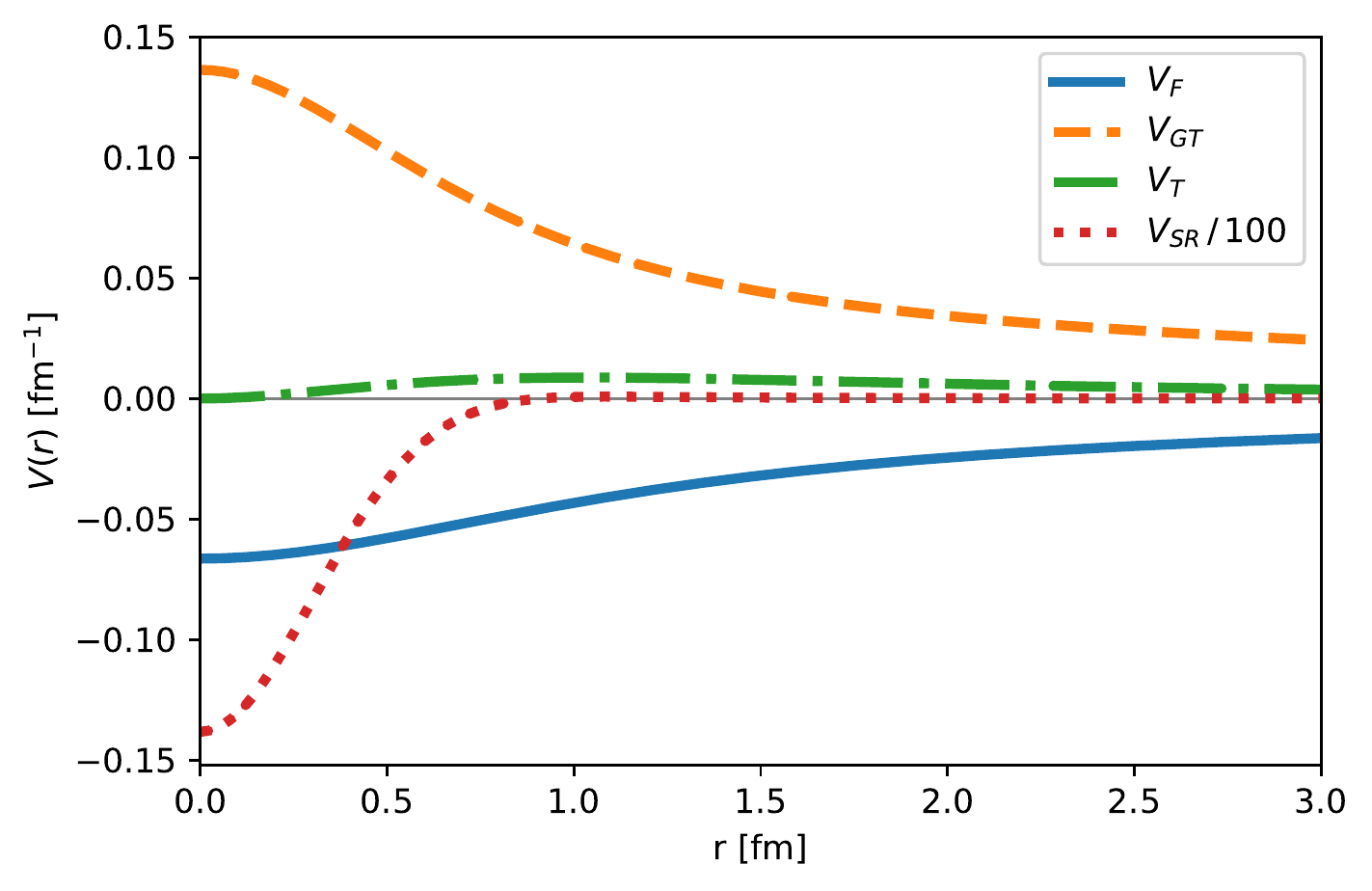}
\caption{Fermi (solid blue curve), Gamow-Teller (dashed orange curve), tensor (dot-dashed green curve), and short-range (dot red curve) transition potentials.\label{fig:pot}}
\end{figure}

Figure~\ref{fig:pot} displays the radial dependence of the transition potentials, and shows that the T component is clearly much smaller than both the F and GT ones. This behavior is reflected in the magnitude of the corresponding NMEs, as highlighted in a number of previous calculations~\cite{Simkovic:1999re,Kortelainen:2007rh,Menendez:2008jp}. Including the form factors regularizes the potentials at short interparticle distances, while the typical $1/r_{ab}$ behavior at large $r_{ab}$ is preserved.

The authors of Ref.~\cite{Cirigliano:2018hja} have demonstrated that an effective field theory approach of the light-neutrino exchange $0\nu\beta\beta$ decay requires a leading-order counter-term to absorb the divergences induced by the long-range neutrino potential and ensure renormalizability. This new short-range (SR) operator is associated with a Fermi spin structure and a SR neutrino potential:
\begin{align}
&O^{0\nu}_S = (4\pi R_A)  \sum_{a \neq b} V^{0\nu}_S(r_{ab}) \tau_a^+ \tau_b^+\,, \\
\label{eq:sr_counter}
&V_S^{0\nu}(r_{ab}) = 2 \frac{g_\nu^{\text{NN}}}{g_A^2} \delta^{(3)}_R(\mathbf{r}_{ab})\,,
\end{align}
where $\delta^{(3)}_R(\mathbf{r}_{ab})$ is a regularized three-dimensional Dirac delta function. In contrast to Ref.~\cite{Cirigliano:2019vdj}, the above definition includes a factor $1/g_A^2$ so that the full light-Majorana transition operator is $O^{0\nu}= O^{0\nu}_F + O^{0\nu}_{GT} + O^{0\nu}_{T} + O^{0\nu}_{S}$. Consistent with the opposite sign in the definition of the F, GT and T contributions, the sign of the SR potential is also different to Ref.~\cite{Cirigliano:2019vdj}.

The value of the new coupling $g_\nu^{\text{NN}}$ arises from non-perturbative QCD dynamics and could in principle be found by matching to lattice-QCD calculations of light-neutrino exchange amplitudes~\cite{Davoudi:2020gxs,Davoudi:2021noh}. It can also be obtained by reproducing the synthetic $2n\rightarrow 2p+2e$ data provided by Refs.~\cite{Cirigliano:2020dmx,Cirigliano:2021qko}. Alternatively, Ref.~\cite{Cirigliano:2019vdj} notes that renormalizing the nucleon-nucleon ($N\!N$) scattering amplitude with Coulomb photon exchange also requires a short-range interaction with coupling $(C_1+C_2)/2$, and connects $0\nu\beta\beta$ and CIB SR couplings: $g_\nu^{\text{NN}} = C_1$. Further, assuming the same value for the two couplings entering the CIB of $N\!N$ potentials, so that $g_\nu^{\text{NN}} \simeq (C_1 + C_2) / 2$, describes well synthetic $2n\rightarrow 2p+2e$ data~\cite{Cirigliano:2021qko}.
We follow this approach, which allows us to evaluate short~range $0\nu\beta\beta$-decay NMEs for a variety of nuclei.

We compute the short-range behavior of nuclear states from the high-quality AV18 $N\!N$ potential. Hence, when evaluating NMEs we make the consistent replacement
\begin{equation}
g_\nu^{\text{NN}} \delta^{(3)}_R(\mathbf{r}_{ab}) \to - \frac{6}{e^2}v_{S1}^{ cd}(r_{ab})\, .
\end{equation}
The full expression for the short-range component of the CIB term of AV18, $v_{S1}^{ cd}(r_{ab})$, for the spin $S=0$, isospin $T=1$ channel can be found in Eq.~(32) of Ref.~\cite{Wiringa:1994wb} and it is displayed in Fig.~\ref{fig:pot}. To better gauge the importance of $V_S^{0\nu}$, we also consider the expression derived from the CIB contribution of the local, $\Delta$-full chiral effective field theory $N\!N$ potential of Ref.~\cite{Piarulli:2016vel}. Specifically, for the NV-Ia$^*$ model $(C_1 + C_2) / 2 = -1.03$ fm$^{2}$ and $\delta^{(3)}_R(\mathbf{r}_{ab}) = e^{-r_{ab}^2 / R_S^2} /(\pi^{3/2} R_S^3)$ with $R_S=0.8$ fm.

Throughout this work, two-body transition densities play a crucial role
\begin{align}
4 \pi r^2 \rho_F(r) &=  \langle \Psi_f | \sum_{a<b}\delta(r-r_{ab})\tau_a^+\tau_b^+|\Psi_i\rangle\,,\nonumber\\
4 \pi r^2 \rho_{GT}(r) &= \langle \Psi_f | \sum_{a<b}\delta(r-r_{ab}) \sigma_{ab}\, \tau_a^+\tau_b^+|\Psi_i\rangle\,,\nonumber\\
4 \pi r^2 \rho_T(r) &= \langle \Psi_f | \sum_{a<b}\delta(r-r_{ab}) S_{ab}\, \tau_a^+\tau_b^+|\Psi_i\rangle\, ,
\label{eq:tran_dens}
\end{align}
and $\rho_S(r) = \rho_F(r)$. All NMEs can be obtained integrating the above densities:
\begin{equation}
M^{0\nu}_\alpha = \int_0^\infty d r\, C^{0\nu}_\alpha(r)\,,
\label{eq:me_int}
\end{equation}
where we define $C^{0\nu}_\alpha(r) \equiv (8 \pi R_A)  4\pi r^2 \rho_\alpha(r) V^{0\nu}_\alpha(r)$ with the additional factor $2$ to compensate the restricted sum $a<b$ in Eq.~\eqref{eq:tran_dens}.

\section{Many-body methods}
\label{sec:many_body}
\subsection{Variational Monte Carlo}
The VMC method solves the Schr\"odinger equation by approximating the true ground state of the system with a suitably parametrized variational wave function $\Psi_T$. The Rayleigh-Ritz variational principle
\begin{align}
\frac{\langle \Psi_T | H | \Psi_T \rangle}{\langle \Psi_T | \Psi_T \rangle} = E_T \geq E_0 
\label{eq:H_exp}
\end{align}
is exploited to find the optimal set of variational parameters. The VMC takes as input the Hamiltonian
\begin{equation}
H = \sum_i -\frac{\hbar^2}{2m} \,{\nabla}_i^2+ \sum_{i<j} v_{ij} + \sum_{i<j<k} V_{ijk} \ ,
\end{equation}
which consists of non-relativistic single-nucleon kinetic energy terms, and two- and three-nucleon interactions. As for the latter, in this work, we utilize the AV18 $N\!N$ interaction~\citep{Wiringa:1994wb} in combination with the UX three-nucleon force.  UX is intermediate between the Urbana and Illinois families of potentials~\cite{Carlson:2014vla}, and is essentially a truncation of the Illinois-7 (IL7) model~\cite{Pieper:2008rui}; it has the form of Eq.(17) in Ref.\cite{Carlson:2014vla}, including two-pion S- and P-wave terms and a short-range isospin-independent repulsion, with the parameter values of IL7, but without the three-pion-ring term or short-range isospin dependence.  The highly successful Green's Function Monte Carlo calculations of light nuclei with AV18+IL7 shown in Ref.\cite{Carlson:2014vla} start from VMC calculations with AV18+UX.

The VMC trial wave function is typically written as
\begin{align}
|\Psi_T \rangle = \Big(1+\sum_{i<j<k} F_{ijk}\Big) \Big(\mathcal{S} \prod_{i<j} F_{ij} \Big) | \Phi_J\rangle\, ,
\label{eq:psi_T}
\end{align}
where $F_{ij}$ and $F_{ijk}$ are two- and three-body correlation operators, respectively and $\mathcal{S}$ denotes a symmetrized product over nucleon pairs. The latter is required for the wave function to be antisymmetric, as, in general, the spin-isospin dependent correlation operators $F_{ij}$ do not commute.  

To account for the alpha-cluster structure of light nuclei the anti-symmetric Jastrow wave function is constructed as a sum over independent-particle terms, $\Phi_A$, each having four nucleons in an $\alpha$-like core and the remaining $(A-4)$ nucleons in $p$-shell orbitals~\cite{Pieper:2002ne}:
\begin{align}
|\Phi_J\rangle&=\mathcal{A}\Bigg[\prod_{i<j<k} f^{c}_{ijk} \prod_{i<j\leq 4} f_{ss}(r_{ij}) \prod_{k\leq 4 < l \leq A} f_{sp}(r_{kl}) \nonumber\\
&\times \sum_{LS[n]} \bigg(\beta_{LS[n]} \prod_{4<l<m\leq A} f_{pp}^{[n]}(r_{lm})\nonumber\\
&\times | \Phi_A(LS[n]J J_z T_z)_{1234:5\dots A}\rangle\bigg)\Bigg]\,.
\label{def:phiJ_gfmc}
\end{align}
The operator $\mathcal A$ denotes an antisymmetric sum over all possible partitions of the $A$ particles into four $s$-shell and $(A-4)$ $p$-shell states. The independent-particle wave function $| \Phi_A(LS[n]J J_z T_z)_{1234:5\dots A}\rangle$ with the desired total angular momentum and projection $J J_z$ values of a given nuclear state is obtained using orbital-spin $LS$ coupling, which is most efficient for nuclei with up to $A\leq12$.  It includes a product over single-particle functions $\phi^{LS}_p(R_{\alpha l})$ ($4<l\leq A$) which are $p$-wave solutions of a particle in an effective $\alpha$-$N$ potential with Woods-Saxon and Coulomb terms.  The symbol $[n]$ is the Young pattern that indicates the spatial symmetry of the angular momentum coupling of the $p$-shell nucleons~\cite{Pudliner:1997ck}. The pair correlation function for particles within the $s$-shell, $f_{ss}$, arises from the structure of the $\alpha$ particle. The $f_{sp}$ is similar to the $f_{ss}$ at short range, but it has a long-range tail that approaches unity at large distances, allowing the wave function to develop a cluster structure, i.e., the asymptotic binding is provided only by the $\phi^{LS}_p(R_{\alpha l})$.  Finally, $f_{pp}$ is set to give the appropriate clustering outside the $\alpha$ core, while $f^c_{ijk}$ is the three-body central correlations induced by the $N\!N$ potential. 

For the $A=6$ case the $^6$He wave function has two spatial symmetry components: $^1$S$_0$[2] and $^3$P$_0$[11], which give a complete p-shell representation.  Since $^6$Be is taken as the charge-symmetric mirror, the $^6$He $\rightarrow$ $^6$Be transition densities are large and have no nodal structure.  For the $A=12$ case, a complete p-shell representation gives the $^{12}$Be wave function two spatial symmetry components: $^1$S$_0$[422] and $^3$P$_0$[332], while $^{12}$C has three additional components, given in Table~\ref{tab:amp12}.
The normalized amplitudes for each component in Table~\ref{tab:amp12} indicate that the dominant states in the initial wave function are a very small part of the final wave function, making the $^{12}$Be $\rightarrow$ $^{12}$C transition densities much smaller than for $A=6$.  They are further reduced by the presence of a node in the transition densities, required by the isospin orthogonality of the two wave functions.

\begin{table}[t]
\begin{center}
\begin{tabular}{c | c c c c c c }
\hline
\hline
       &$^1$S$_0$[44] &$^3$P$_0$[431] &$^1$S$_0$[422] &$^5$D$_0$[422] &$^3$P$_0$[332] \\
\hline
$^{12}$Be &  -    &  -    & 0.983 &  -    & 0.186 \\
$^{12}$C  & 0.947 & 0.314 & 0.055 & 0.015 & 0.033 \\
\hline 
\end{tabular}
\caption{Normalized amplitudes for different spatial symmetry components in the VMC wave functions for $^{12}$Be and $^{12}$C. 
\label{tab:amp12}}
\end{center}
\end{table}

In an earlier comparison of VMC and SM calculations~\cite{Wang:2019hjy} the VMC wave function for $^{12}$C included only the leading [44] and [431] components.  While these are by far the dominant part of the final state wave function, the small [422] and [332] components included here have a much greater spatial overlap with the $^{12}$Be wave function, leading to a significant change, particularly at long range, in the transition densities.  In particular, the GT NME with the more complete $^{12}$C wave function is about double that of the previous VMC calculation, while the F and T increase by 10-20\%.
This improves the agreement between VMC and the earlier SM calculations of Ref.\cite{Wang:2019hjy}.

The expectation values of quantum mechanical operators of the form of Eq.(\ref{eq:H_exp}) contain multi-dimensional integrals over all nucleon positions
\begin{align}
\frac{\langle \Psi_T| \mathcal O |\Psi_T \rangle}{\langle \Psi_T |\Psi_T \rangle} = \frac{\int d\mathbf{R} \Psi^\dagger_T(\mathbf{R}) \mathcal O \Psi_T(\mathbf{R}) }{\int d\mathbf{R}\Psi^\dagger_T(\mathbf{R}) \Psi_T(\mathbf{R})}\,,
\label{eq:expectation}
\end{align}
and Metropolis-Hastings Monte Carlo techniques are employed to efficiently evaluate them. These Monte Carlo samples are also used to compute the two-body transition densities of Eq.\eqref{eq:tran_dens} and to estimate their statistical uncertainties.

\subsection{Nuclear Shell Model}

The nuclear SM is one the most successful nuclear many-body methods to describe the properties of ground and excited states, including electromagnetic and weak transitions~\cite{Brown:1988vm,Caurier:2004gf,Otsuka:2018bqq}. It is also one of the common methods to compute $0\nu\beta\beta$-decay NMEs~\cite{Menendez:2008jp,Horoi:2015tkc,Iwata:2016cxn,Coraggio:2020hwx}.

In order to handle both light and heavy nuclei, the nuclear SM simplifies the many-body problem by restricting to a relatively small configuration space consisting of one or two harmonic oscillator shells. This excludes from the calculation the core---filled with nucleons---below and the high-energy orbitals---empty---above the configuration space, but their impact is captured by an effective interaction corresponding to the configuration space. The resulting many-body Schr\"odinger equation is
\begin{equation}
\label{eq:sm_diag}
H_{\text{eff}}\,| \Psi_{\text{SM}} \rangle=E\,|\Psi_{\text{SM}}\rangle,
\end{equation}
which we solve using the $J$-coupled code NATHAN~\cite{Caurier:2004gf}.
Even though {\it ab initio} approaches allow one to obtain effective interactions solely based on $N\!N$ and three-nucleon forces~\cite{Stroberg:2019mxo},
in this work we use high-quality interactions obtained from $N\!N$ potentials complemented with small phenomenological adjustments, mostly on the monopole part~\cite{Caurier:2004gf}.

For light $A\leq12$ nuclei we use the p- and sd-shell configuration space and the PSDMWK interaction~\cite{Warburton:1992qu,Warburton:1992rh} corrected for center-of-mass contamination. In heavier nuclei we use the same configuration space and SM interactions as in previous SM studies~\cite{Menendez:2008jp,Menendez:2017fdf}: the pf-shell with the KB3G interaction for $^{48}$Ca, the $1p_{3/2}$, $1p_{1/2}$, $0f_{5/2}$, $0g_{9/2}$ space with the GCN2850 interaction~\cite{Menendez:2008jp} for $^{76}$Ge, and the $1d_{5/2}$, $2s_{1/2}$, $1d_{3/2}$, $0g_{7/2}$, $0h_{11/2}$ space with the GCN5082 interaction~\cite{Menendez:2008jp} for $^{130}$Te and $^{136}$Xe.

The SM wave functions from Eq.~\eqref{eq:sm_diag} directly provide energies and other observables not dependent on radial degrees of freedom. However, for $0\nu\beta\beta$ decay the spatial part is relevant as well, and usually a harmonic oscillator (HO) basis is used for single-particle states~\cite{Menendez:2008jp,Menendez:2017fdf}.
Here we follow the improved approach of Ref.~\cite{Wang:2019hjy} and obtain our transition densities replacing the standard HO spatial single-particle states by Woods-Saxon (WS) ones, which reflect a more realistic long-range asymptotic behavior. We consider two kinds of WS potential: first, the standard parametrization from Suhonen (WSS)~\cite{Suhonen}; second, the potential proposed by Ref.~\cite{Wang:2019hjy} adjusted to the experimental neutron and proton separation energies and taking all orbitals in the configuration space as bound (WSW)---however, these conditions cannot be met for $A=6$ which we only study with WSS. We have checked that alternative WS parametrizations~\cite{Schwierz:2007ve} give very similar results to WSS.

\begin{figure}[t]
\includegraphics[width=\linewidth]{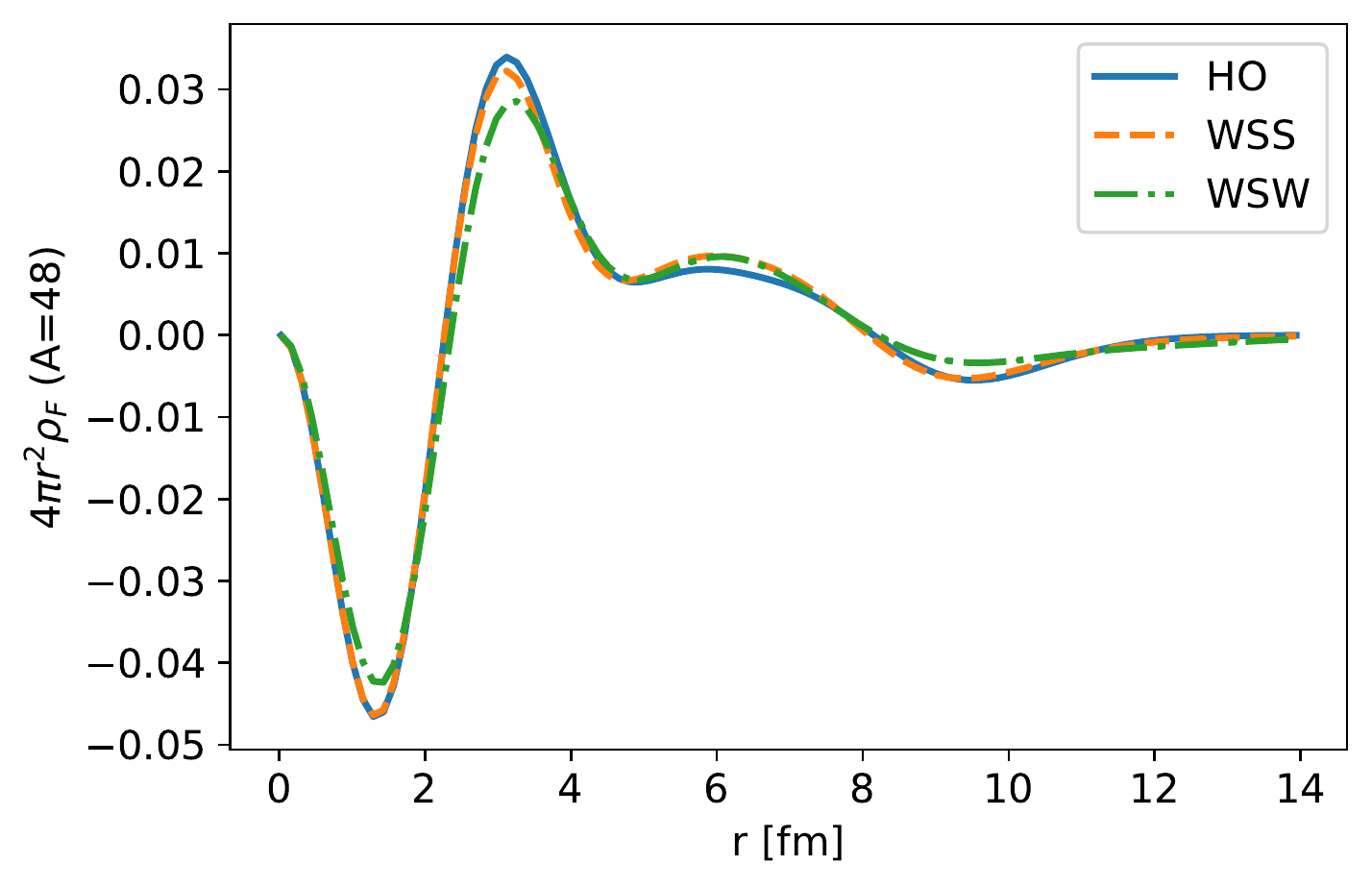}
\caption{Fermi transition density for $A=48$ using the SM with HO, WSS, and WSW single-particle orbitals.~\label{fig:rho_F_A48_HO_WS}}
\end{figure}

In light nuclei, Ref.~\cite{Wang:2019hjy} shows that SM results with WSW orbitals greatly improve the agreement with VMC ones. The improvement with WSS is similar.
However, extending the SM results with WS single-particle orbitals to heavy nuclei is challenging, and only HO calculations are currently feasible. Fortunately, we have tested in $A=48$ that the differences between using HO and WS orbitals become smaller for heavier systems. Figure~\ref{fig:rho_F_A48_HO_WS} shows minor differences between the F transition density computed employing the two different parametrizations of the WS orbitals and the HO one for single-particle orbitals. Likewise, the GT and T transition densities are also very similar. 

Since the nuclear SM deals with regularized effective interactions, part of the short-range dynamics is missing in the wave functions. This shortcoming is common to other non {\it ab initio} approaches such as the QRPA, energy-density functional theory and interacting boson model. Because the short-range dynamics can impact $0\nu\beta\beta$ decay NMEs, typically calculations correct for missing SRCs via a Jastrow-type function. Different parametrizations have been provided by Miller-Spencer~\cite{Miller_Spencer_1976} or based on Argonne and CD-Bonn potentials~\cite{Simkovic2009}, or by others \cite{CRUZTORRES2018304,Benhar:2014cka,Roth2005UCOM}.
However, in this work we do not include any additional correlations of this kind, as we introduce the correct short-range dynamics captured by the VMC calculations using the GCF.

\subsection{Generalized Contact Formalism}

The GCF is an effective theory
for describing the impact of SRCs
on a variety of nuclear distributions and observables. This formalism has proven extremely successful in modeling the short-range and high-momentum parts of different nuclear densities~\cite{Weiss:2015mba,Weiss:2016obx,Cruz-Torres2020},
and also large momentum transfer electron scattering experiments sensitive to SRCs~\cite{Weiss:2018tbu,schmidt20,Pybus:2020itv,Duer:2018sxh,weiss2020inclusive} and other reactions~\cite{Weiss:2014gua,Weiss_EPJA16,Weiss_2016,WEISS2019484}. In a high-resolution picture, when two or more particles are close to each other, and, therefore, strongly interacting, the SM solution for the wave function---based on a regularized potential---becomes inaccurate~\cite{Hen:2013oha}. For example, SRCs lead to a significant occupation of high momentum states absent in the SM. On the other hand, QMC methods fully capture these features but are limited to light nuclei. Therefore, the GCF provides an ideal framework to quantitatively incorporate the correct short-range behavior into shell-model calculations of heavy nuclei. 

The GCF is based on scale separation, leading to wave-function factorization
when two particles are very close to each other. Explicitly, any nuclear wave
function $\Psi(\bs{r}_1,\bs{r}_2,...,\bs{r}_A)$ is expected to obey the asymptotic form~\cite{Weiss:2015mba}
\be
\Psi \xrightarrow[r_{ij}\rightarrow 0]{}
\sum_\alpha \varphi^\alpha(\bs{r}_{ij}) A^\alpha(\bs{R}_{ij},\{\bs{r}_k\}_{k\neq i,j}).
\ee
Here $\bs{r}_{ij}=\bs{r}_j-\bs{r}_i$ and $\bs{R}_{ij}=(\bs{r}_i+\bs{r}_j)/2$ are
the relative and center of mass coordinates of the pair, and
$\alpha$ denotes its quantum numbers, i.e., parity $\pi_\alpha$, spin $s_\alpha$,
total angular momentum $j_\alpha$, and projection $j_{\alpha z}$, and total isospin $t_\alpha$
and projection $t_{\alpha z}$. Isospin quantum numbers are relevant to keep the nuclear wave function $\Psi$ anti-symmetric under permutations of any two nucleons. This convention is equivalent to the one in most previous GCF studies considering wave functions anti-symmetric under separate permutations of protons and neutrons. The solution of the zero-energy two-body Schr\"odinger equation $\varphi^\alpha(\bs{r}_{ij})$ describes the pair dynamics when the two nucleons are close together. It is a universal function, i.e. identical for all nuclei and all quantum states, but depends on the model and the nuclear
interaction. It can be written as
\be
\varphi^{\alpha}\left(\bs{r}\right)\equiv
\eta_{t_\alpha,t_{\alpha z}} \sum_{\ell_\alpha  \in \pi_\alpha } \phi^{\alpha}(r)
\left[Y_{\ell_\alpha}\left(\hat{\bs{r}}\right)\otimes\chi_{s_\alpha }\right]^{j_\alpha m_\alpha }
\;,
\ee
where $\eta_{t,t_z}$ is the isospin function, $Y_{lm}$
spherical harmonics, $\chi_{sm}$ the two-body spin function, and the sum runs over orbital angular momenta $\ell_\alpha$ of correct parity $\pi_\alpha$ that can couple with $s_\alpha$ yielding $j_\alpha$. The radial dependence is modeled by $\phi^{\alpha}(r)$, which is independent of $j_{\alpha z}$ and, to good accuracy, also of $t_{\alpha z}$ due to isospin symmetry. 

Based on this asymptotic form, the nuclear {\it contacts}
for a nucleus with $A$ nucleons are defined
as
\be
C^{\alpha\beta}=\frac{A(A-1)}{2}\langle A^\alpha | A^\beta \rangle.
\label{eq:contact}
\ee
The factor $A(A-1)/2$ appears in place of the number of proton-proton, neutron-proton or neutron-neutron pairs present in previous publications because here the wave function is anti-symmetric under permutation of any two nucleons. 

The diagonal contacts $C^{\alpha\alpha}$ are proportional to the number of correlated pairs in the nucleus with quantum numbers $\alpha$. However, in this work we apply the GCF to describe the short-range behavior of the two-body densities relevant for $0\nu\beta\beta$ transitions. Hence, we define new contact parameters
that involve different initial ($i$) and final ($f$) nuclear states as 
\be
C^{\alpha\beta}(f,i)=\frac{A(A-1)}{2}\langle A^\alpha(f)|A^\beta(i) \rangle.
\label{eq:C_fi}
\ee
Using the above definition, we can write the dominant
contributions to the transition densities defined in
Eq.~\eqref{eq:tran_dens} at short distances.
For F and GT transitions, we expect pairs in an s-wave state, mainly with $s=0$, $j=0$,
$t=1$. Denoting the corresponding contact parameter for a transition of two neutrons to two protons ($nn\rightarrow pp$) with such quantum numbers as $C_{pp,nn}^0(f,i)$, the F transition density can be expressed as
\be
\rho_F(r) \xrightarrow[r\rightarrow 0]{} 
\frac{1}{4\pi}|\phi^0(r)|^2 C_{pp,nn}^0(f,i),
\label{eq:rho_F_GCF}
\ee
where $\phi^0(r)$ is the radial function for the $\ell=0$, $s=0$, $j=0$, $t=1$ channel. Since $\rho_S=\rho_F$, the above asymptotic form is valid for transition density associated with the short-range operator of Eq.~\eqref{eq:sr_counter}. 
As for the GT transition, the $\sigma_{ab}$ operator leads to 
a factor of $(-3)$ in this $s=0$ channel and we similarly obtain
\be
\rho_{GT}(r) \xrightarrow[r\rightarrow 0]{} 
-\frac{3}{4\pi} |\phi^0(r)|^2 C_{pp,nn}^0(f,i),
\label{eq:rho_GT_GCF}
\ee
which implies the following relation between the F and GT densities for short distances
\be
\rho_{GT}(r \lesssim 1\;\textrm{fm}) =
-3 \rho_F (r \lesssim 1\; \textrm{fm}).
\ee
Based on our previous experience with two-body densities~\cite{Weiss:2016obx,Cruz-Torres2020},
these expressions should provide a good description of the transition densities for $r \lesssim 1$ fm.

To calculate the $0\nu\beta\beta$ matrix elements, we wish to combine the GCF expressions, valid at short distances, and the long-range behavior of the nuclear SM many-body wave functions. The main unknowns in this approach are the values of the relevant
contacts, which in general depend upon the nucleus and on the particular nuclear interaction. Nevertheless, previous studies have shown that for the case of the contacts defined in Eq.~\eqref{eq:contact}, contact ratios $C^{\alpha\alpha}(X)/C^{\alpha\alpha}(Y)$,
for any two nuclei $X$ and $Y$, are model independent~\cite{Cruz-Torres2020,WEISS2019484,PhysRevC.104.034311}. In this sense, contact ratios can be interpreted as long-range, low-resolution quantities that do not depend on the details of the nuclear interaction. 

Such a model independence is expected to hold also for ratios of the contacts defined in Eq.~\eqref{eq:C_fi}.
Therefore, the ratio of the contacts $C^{0,\text{SM}}_{pp,nn}(f_1,i_1)/C^{0,\text{SM}}_{pp,nn}(f_2,i_2)$---indices $1$ and $2$ denote different $0\nu\beta\beta$ decays---is inferred from SM transition densities at short distances. Then, the contact $C^{0,V_{N}}_{pp,nn}(f_2,i_2)$ is obtained by fitting the short-range behavior to the transition density ``$2$'' computed with  VMC for a given realistic nuclear interaction $V_N$. Finally, the contact for transition ``1'' is obtained exploiting the 
model independence of the ratios:
\be
C^{0,V_{N}}_{pp,nn}(f_1,i_1) = \frac{C^{0,\text{SM}}_{pp,nn}(f_1,i_1)}{C^{0,\text{SM}}_{pp,nn}(f_2,i_2)}
C^{0,V_{N}}_{pp,nn}(f_2,i_2).
\label{eq:ratio_sm_vmc}
\ee
This procedure allows us to obtain contact values
of heavy nuclei for any nuclear interaction, using only
a single {\it ab-initio} calculation for light nuclei and SM ones for both heavy and light nuclei. In Sec.~\ref{sec:results} we provide ample evidence on the accuracy of the model independence of contact ratios.

The contact value $C^{0,V_{N}}_{pp,nn}(f_1,i_1)$ and the corresponding short-range radial function fully determine the short-range part ($r \lesssim 1$ fm) of the transition densities for a given nuclear interaction. On the other hand, the SM is expected to provide high-quality transition densities at long distances. Thus, we merge the GCF and SM results continuously, by scaling the SM transition densities to match the GCF expression around $r\simeq1$~fm. This approach, dubbed GCF-SM, allows us to obtain the F and GT transition densities for any given nuclear interaction---including heavy nuclei where direct {\it ab initio} calculations with high-resolution potentials are currently not possible. We integrate the resulting transition densities as in Eq.~\eqref{eq:me_int} to evaluate the relevant $0\nu\beta\beta$ NMEs.

In the case of the T transitions
the leading contribution is expected to come from p-wave channels. There are three such channels (with $j=0,1,2$) which complicates the analysis. In addition, comparing SM and VMC calculations, it seems that the model-independence of the ratios does not hold for the T case. For this reason, in this work we estimate the T matrix element by the SM results with a $50\%$ uncertainty. This should not have a big impact on the total NME as the T part is expected to be small compared to the GT contribution~\cite{Menendez:2008jp,Engel:2016xgb}.

\begin{figure}[t]
\includegraphics[width=\linewidth]{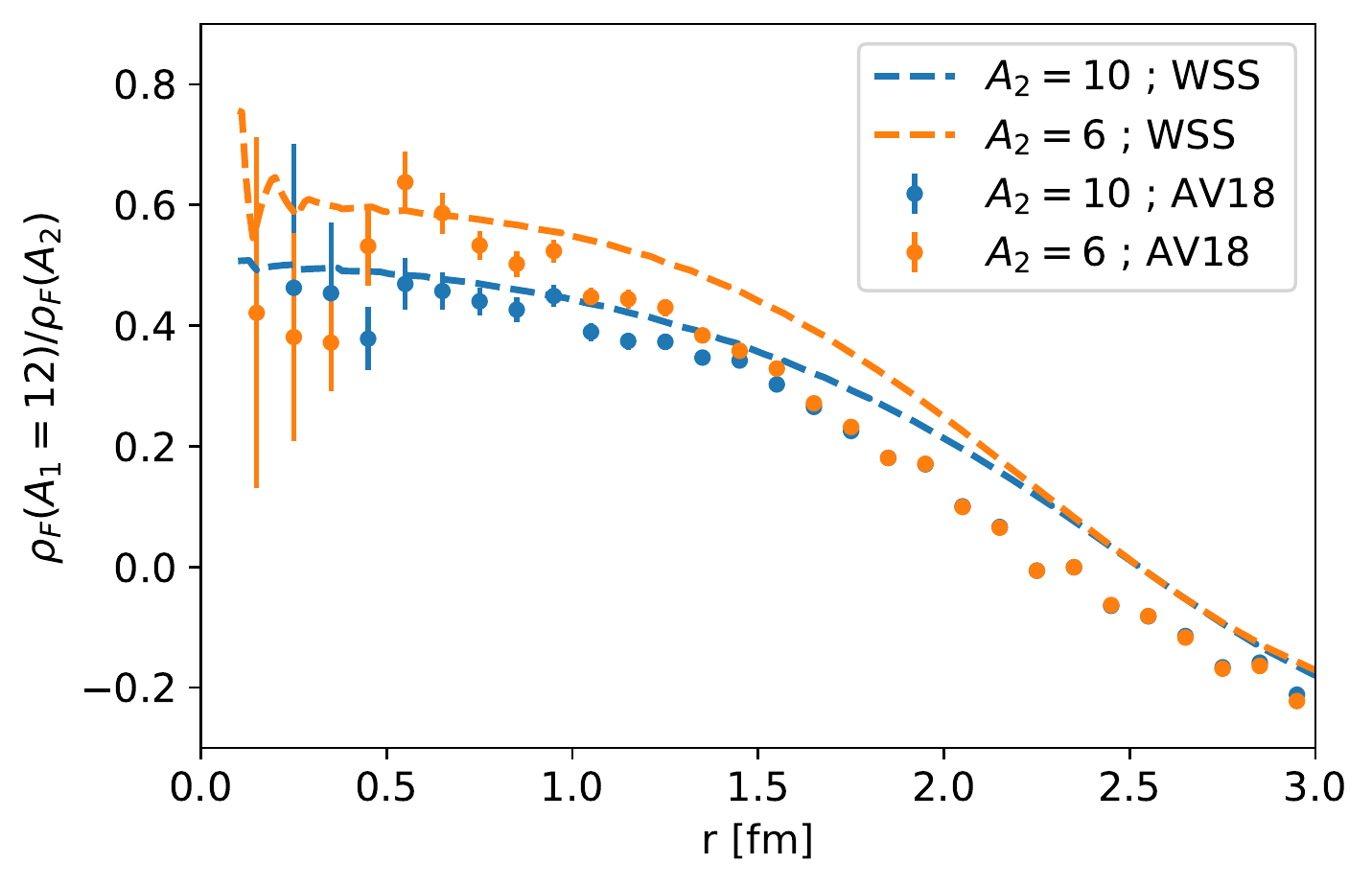}
\caption{Fermi transition density ratio for different
nuclei, calculated using different interactions (AV18 and SM). At short distances, the ratio plateaus and is similar for the two interaction models.~\label{fig:model_ind}}
\end{figure}

\section{Results and discussion}
\label{sec:results}
\subsection{Light nuclei}
\label{sec:light}
In order to use the GCF to describe the short-range part of the transition densities, we evaluate the contact $C_{pp,nn}^0(f,i)$ assuming the model independence of contact ratios. In light nuclei, the availability of both VMC and SM $0\nu\beta\beta$ transition densities allows us to test the accuracy of this approach. Figure~\ref{fig:model_ind} displays two ratios of F transition densities: $^{12}\textrm{Be} \to\, ^{12}\textrm{C}$ decay over $^{10}\textrm{Be} \to\, ^{10}\textrm{C}$ and over $^{6}\textrm{He} \to\, ^{6}\textrm{Be}$, obtained with both the VMC and SM. At short distances the ratio reaches a plateau, as expected according to Eq. \eqref{eq:rho_F_GCF}, representing the contact ratio of the two transitions. In addition, in both cases (e.g. $A=10$ and $A=6$) the value of the ratio at short distances is very similar for the VMC and SM calculations, indicating model independence.

\begin{figure}[t]
\includegraphics[width=\linewidth]{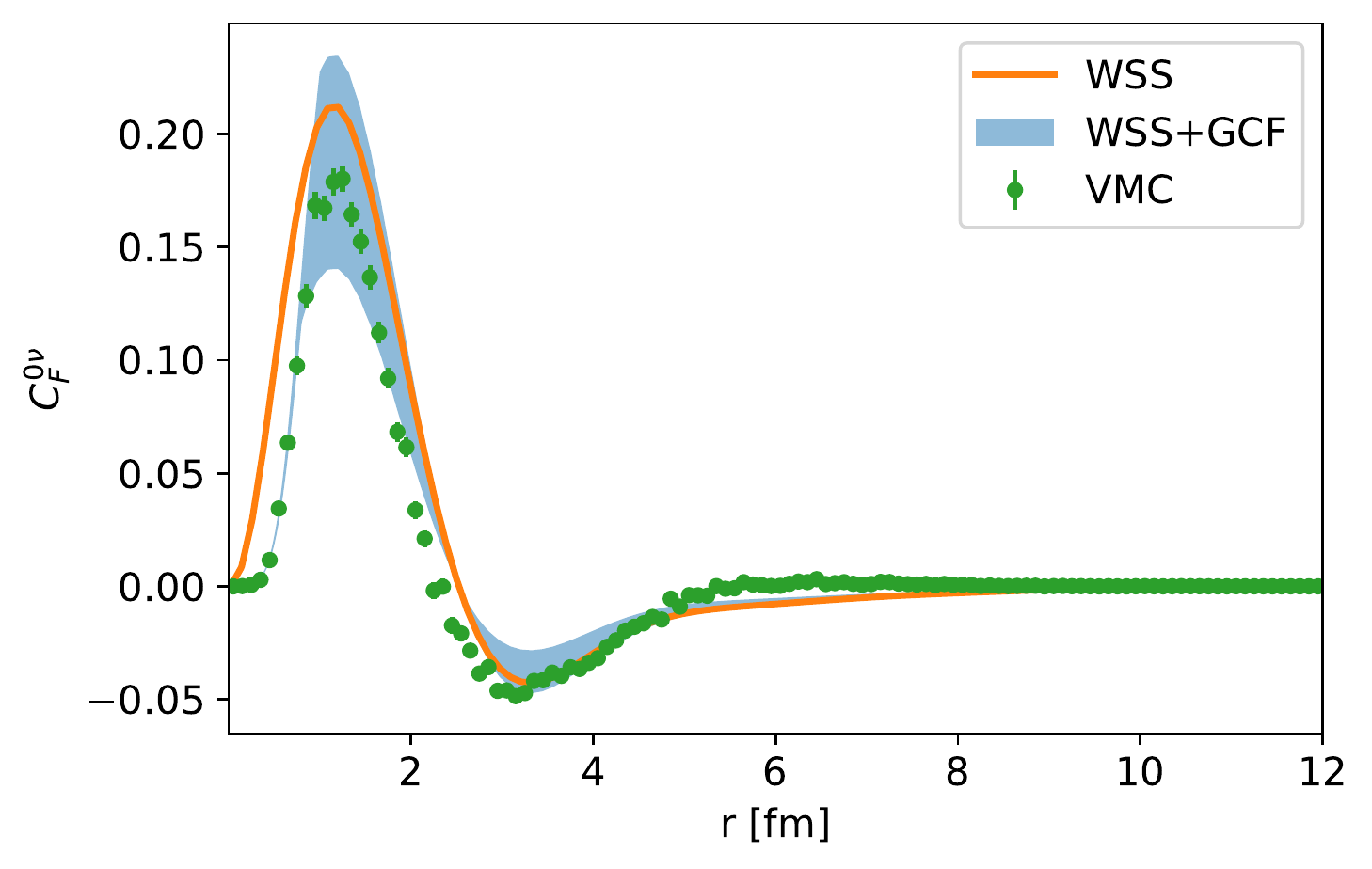}
\includegraphics[width=\linewidth]{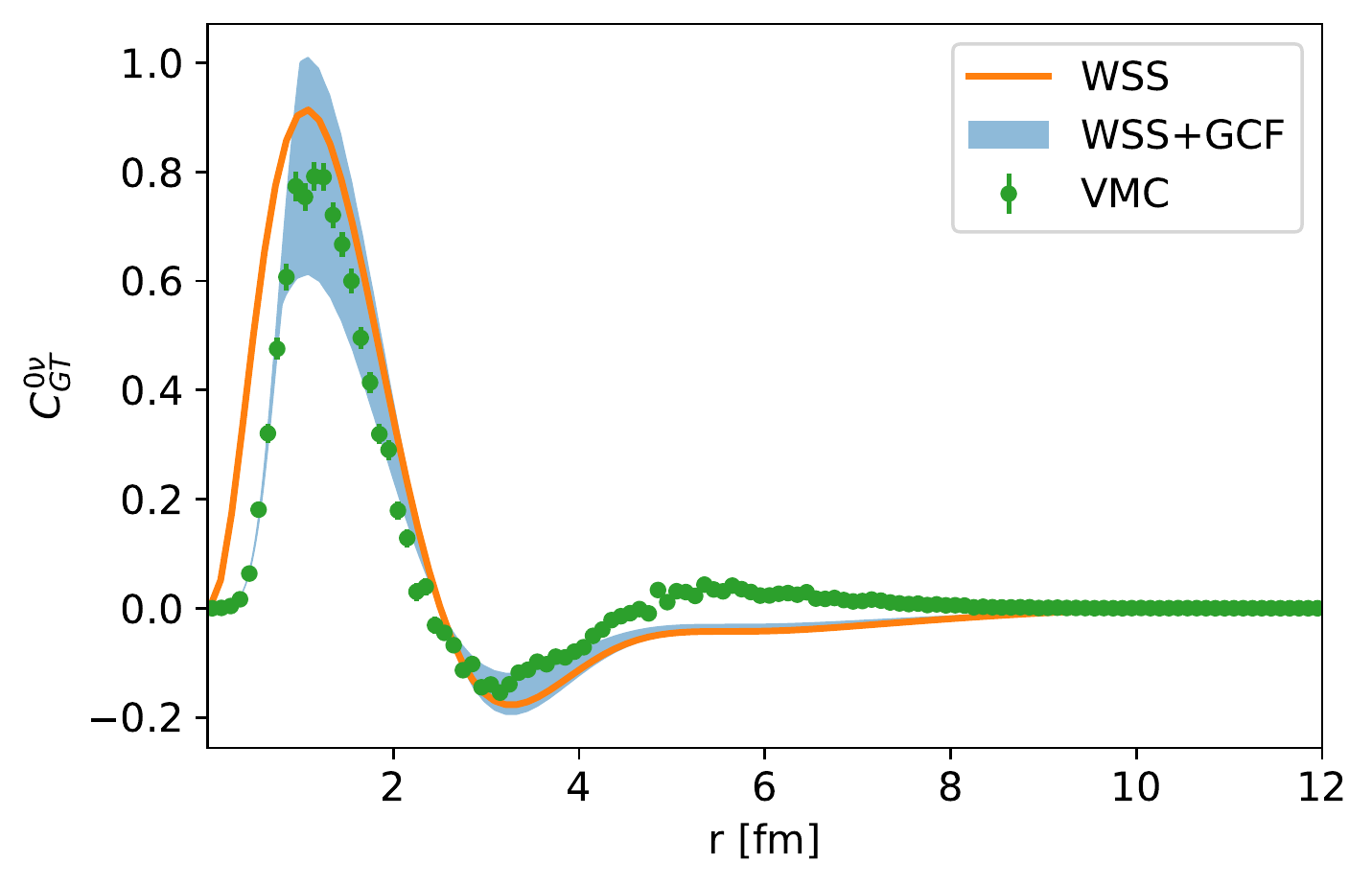}
\includegraphics[width=\linewidth]{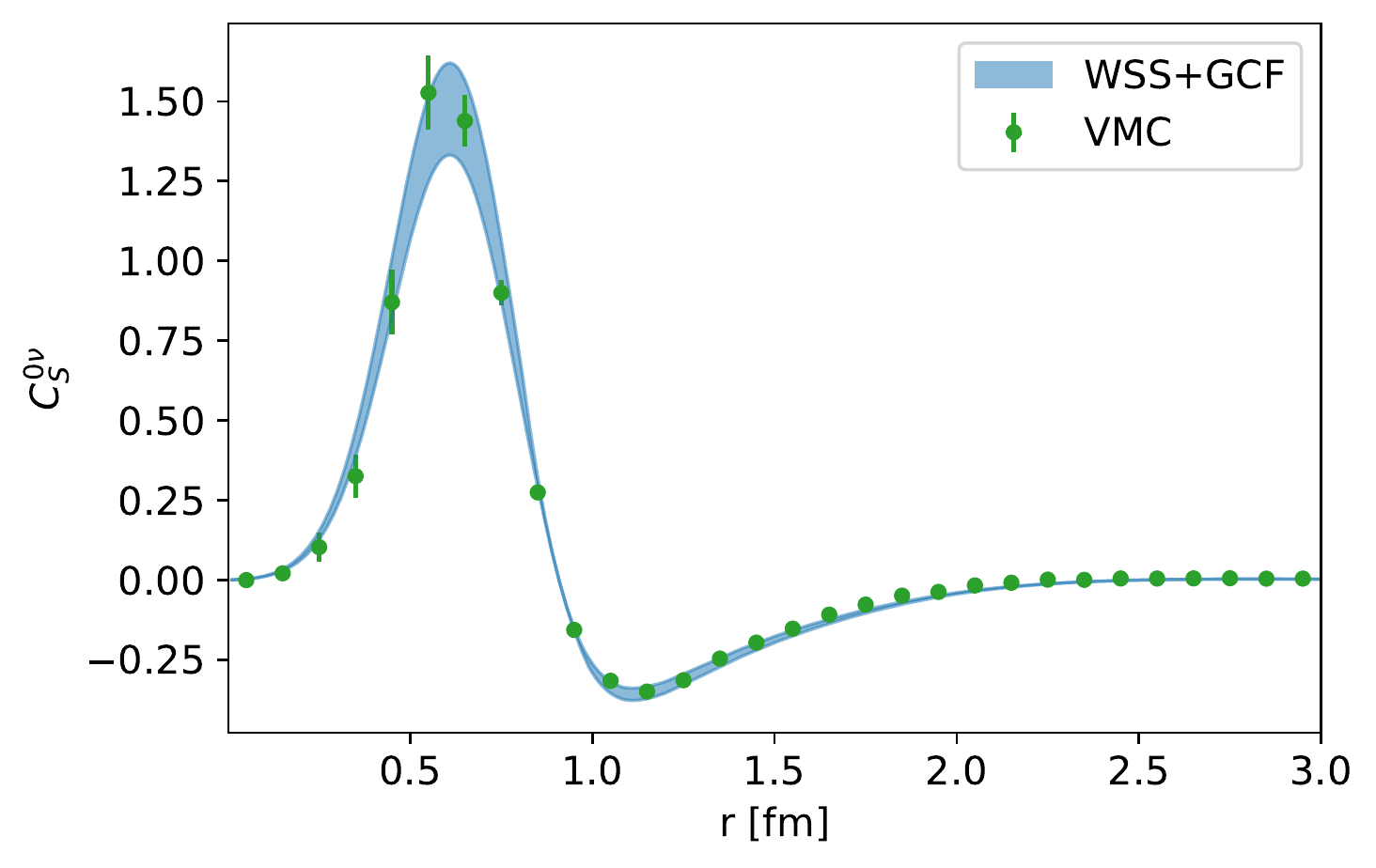}
\caption{Fermi, Gamow-Teller and short-range transition densities calculated with the SM WSS (orange line), VMC (green points), and the combination of the GCF and SM WSS (blue band) approaches. For the latter, only VMC calculations for $A=6$ and $A=10$ are used to extract the $^{12}\textrm{Be}\rightarrow ^{12}\textrm{C}$ contact value.~\label{fig:A12}}
\end{figure}

To gauge the accuracy of our approach, we use the VMC transitions for $A=6$ and $A=10$ to predict results for $A=12$. The contact $\textrm{C}_{pp,nn}^0(^{10}\textrm{Be},^{10}\textrm{C})$ is fitted to
the VMC calculations using the functional form of Eq.~\eqref{eq:rho_F_GCF}. We then obtain the $A=12$ contact $\textrm{C}_{pp,nn}^0(^{12}\textrm{Be},^{12}\textrm{C})$, based on the model independence of Eq.~\eqref{eq:ratio_sm_vmc}, multiplying $\textrm{C}_{pp,nn}^0(^{10}\textrm{Be},^{10}\textrm{C})$ by the average value of the SM ratio $\rho_F(A=12)/\rho_F(A=10)$ at short distances. We repeat the same procedure using as input the VMC transition densities for $A=6$. Comparing the two approaches yields $\lesssim 10\%$ differences in the extracted contact values, and we take the average as our best estimate for $\textrm{C}_{pp,nn}^0(^{12}\textrm{Be},^{12}\textrm{C})$, associated with a $10\%$ uncertainty.

Once the value of the contact is determined, we construct the short-range part of the F, GT and SR transition densities with Eqs.~\eqref{eq:rho_F_GCF} and \eqref{eq:rho_GT_GCF}. We highlight the consistency of the approach, as the value of the contact is extracted from VMC calculations using the AV18+UX Hamiltonian, and the two-body function $\phi^0(r)$ is consistently computed with the AV18 $N\!N$ force.

For long distances we use the SM transition densities, after re-scaling them so that they are continuously merged with the short-range part. This part is expected to be well described by the SM, given the very good description it gives of nuclear structure and spectroscopy~\cite{Caurier:2004gf,Otsuka:2018bqq}.
In particular, long-range correlations have been studied extensively in $0\nu\beta\beta$ studies~\cite{Hinohara:2014lqa,Menendez:2014ena,Yao:2021wst}, and the SM has proven to capture well important correlations such as those related to high-seniority components~\cite{Caurier:2007wq} or proton-neutron pairing~\cite{Menendez:2015kxa}. Therefore, we expect that the GCF-SM describes also well the long-range part of the transition densities.

The upper, middle, and lower panels of Fig.~\ref{fig:A12} display the F, GT and SR transition densities obtained with the above procedure. The band is obtained by randomly varying separately the contact value within its $10\%$ uncertainty, and the matching point of GCF and SM between $0.8-1.0$ fm.  
Figure~\ref{fig:A12} shows an overall good agreement between the GCF-SM and the VMC results. In particular, our new method greatly improves upon the short-range behavior of the SM transitions densities, bringing them in excellent agreement with the VMC ones. It has to be noted that while introducing ad-hoc Jastrow-like correlations into SM calculations certainly ameliorates their short-range behavior, the agreement with {\it ab-initio} results is not as good~\cite{Wang:2019hjy}. In addition, the GCF can readily accommodate different interactions and correct SM calculations in a consistent fashion. The long range part of the transitions ($r\gtrsim 1$ fm), taken from the SM, also agrees well with the VMC. In this regard, an important role is played by the complete p-shell representation of the VMC $^{12}$C wave function utilized in this work, in contrast to earlier VMC calculations~\cite{Pastore:2017ofx,Cirigliano:2019vdj} that only included the leading [44] and [431] components. As expected, the transition density of the SR operator is almost perfectly described by our approach, at both short and long distances. 

\begin{figure}[t]
\includegraphics[width=\linewidth]{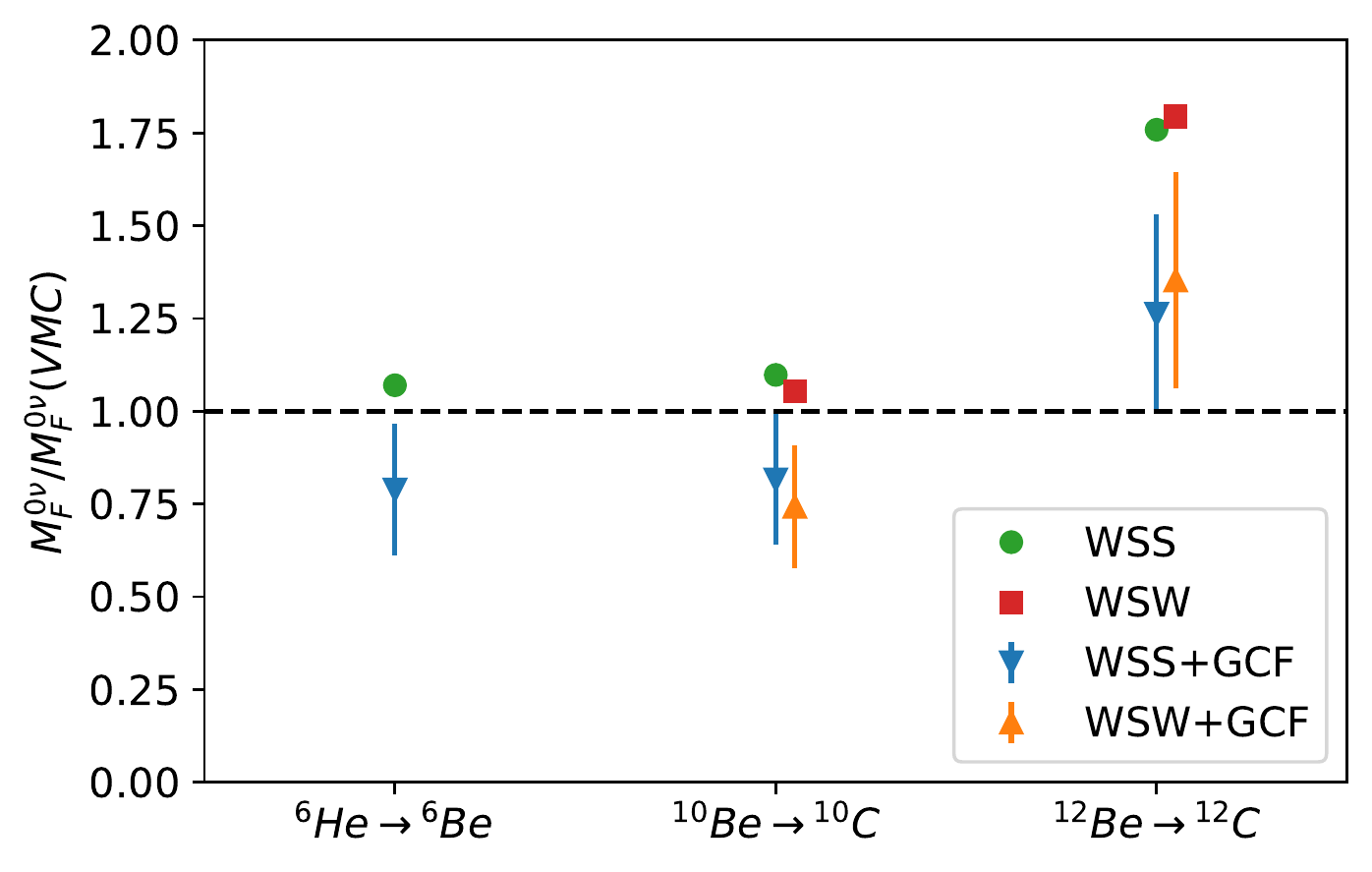}
\includegraphics[width=\linewidth]{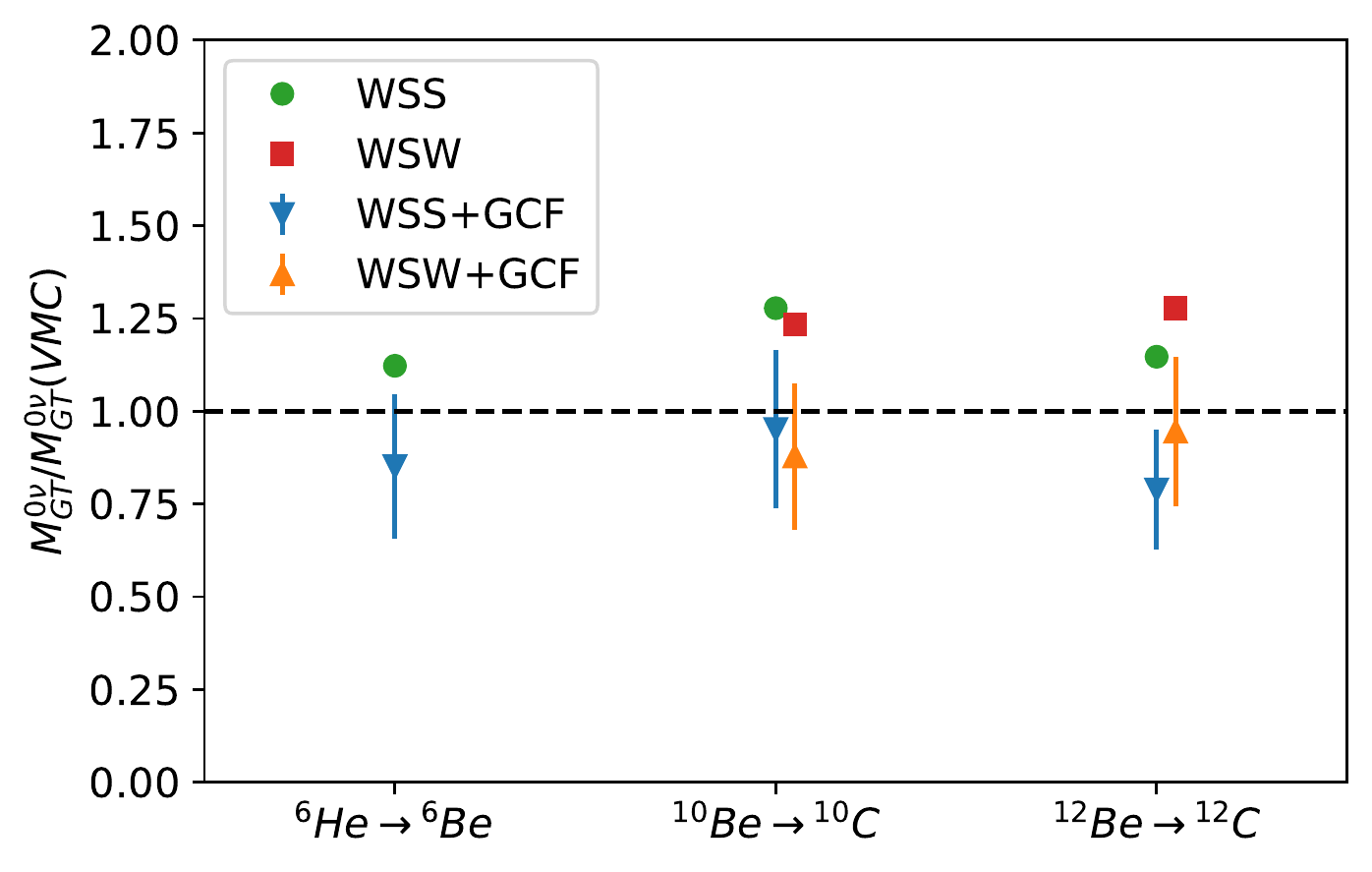}
\includegraphics[width=\linewidth]{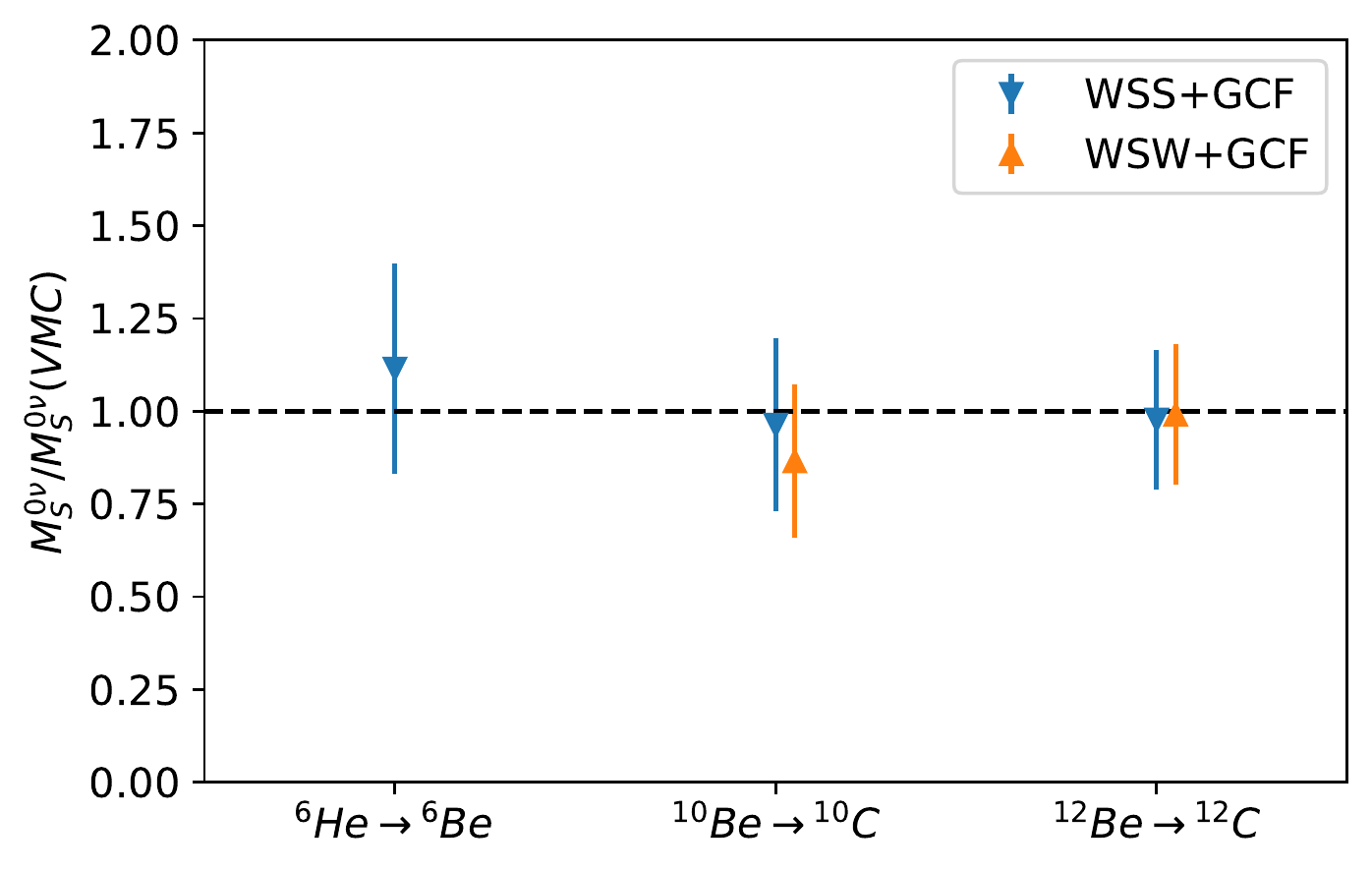}

\caption{Ratio of nuclear matrix elements of the F (upper panel), GT (central panel), and SR (lower panel) operators. Results obtained combining the GCF and SM, or just the SM approach, divided by the VMC ones with the AV18+UX Hamiltonian. Both WSS and WSW parametrizations are used for the SM calculations. ~\label{fig:ME_light}}
\end{figure}

We further gauge the accuracy of the GCF-SM method by using either both the $A=10$ and $A=12$, or the $A=6$ and $A=12$ VMC transition densities, to predict $A=6$, or $A=10$ NMEs $M_\alpha^{0\nu}$, respectively, using Eq.~\eqref{eq:me_int}. Figure~\ref{fig:ME_light} compares the GFC-SM the $A=6$, $A=10$, and $A=12$ F (upper panel) GT (middle panel) and SR (lower panel) NMEs for these nuclei to the VMC and standard SM results---see Table \ref{tab:MEs_light} for their numerical values. The VMC and GCF-SM matrix elements are consistent within error bars---little dependence on the particular WS parametrization is observed---and SR NMEs agree especially well. The SR SM values without correcting for SRC, not shown in Fig. \ref{fig:ME_light}, are about 10 times larger than the VMC ones, due to the scale of the AV18 SR potential, see Fig.~\ref{fig:pot}. In almost all cases, complementing SM calculations with the GCF improves over the SM results.
Perhaps more importantly, while uncertainties of the SM calculations are difficult to estimate, the GCF-SM error estimates appear to be robust. 

We note that using SR transition potential from the NV-Ia* interaction changes the value of $M_S^{0\nu}$ by less than $20\%$, despite the differences in the shape of the transition densities $C_S^{0\nu}(r)$ are more dramatic. Nonetheless, fully consistent calculations of SR operators for potentials other than AV18 require using the appropriate two body function $\phi^0(r)$ and the corresponding VMC transition densities. This is left for future work.

\begin{table}[t]
\begin{center}
\begin{tabular}{c | c | c c c c c  }
\hline
\hline
 Transition & Method & F       & GT      & SR  \\
\hline
\multirow{3}{*}{$^{6}$He $\rightarrow$ $^{6}$Be} &
VMC               & 0.935    & 3.706     & 0.296  \\
&WSS              & 1.001    & 4.160     & 3.286  \\
&WSS+GCF          & 0.74(17) & 3.16(72)  & 0.33(8)\\
\hline 
\multirow{5}{*}{$^{10}$Be $\rightarrow$ $^{10}$C} &
VMC               & 1.182    & 3.647     & 0.513  \\
&WSW              & 1.246    & 4.500     & 4.457  \\
&WSS              & 1.297    & 4.660     & 4.721  \\
&WSW+GCF          & 0.88(20) & 3.20(72)  & 0.44(11) \\
&WSS+GCF          & 0.96(21) & 3.47(78)  & 0.49(12) \\
\hline 
\multirow{6}{*}{$^{12}$Be $\rightarrow$ $^{12}$C} &
VMC$_2$           & 0.102    & 0.365     & 0.347   \\
&VMC$_5$          & 0.113    & 0.757     & 0.348   \\
&WSW              & 0.203    & 0.968     & 2.531   \\
&WSS              & 0.199    & 0.868     & 2.562   \\
&WSW+GCF          & 0.15(3)  & 0.72(15)  & 0.35(7) \\
&WSS+GCF          & 0.14(3)  & 0.60(12)  & 0.34(7) \\

\end{tabular}
\caption{Fermi, Gamow-Teller, and short-range NMEs for the $^{6}$He $\rightarrow$ $^{6}$Be, $^{10}$Be$\rightarrow ^{10}$C and $^{12}$Be $\rightarrow$ $^{12}$C transitions calculated using different methods. VMC$_2$ stands for VMC calculations with 2 components for $^{12}$C, while VMC$_5$ includes 5 components. The latter is used to extract the contact ratios.
\label{tab:MEs_light}}
\end{center}

\end{table}

\subsection{Heavy nuclei}\label{sec:heavy}
Having validated the accuracy of the GCF-SM approach on light nuclei, we now turn our analysis to nuclei relevant to the $0\nu\beta\beta$ decay experimental program, currently beyond the reach of the VMC method. We obtain GCF-SM predictions of $0\nu\beta\beta$ transitions analogously to light nuclei, the only difference being that we use all the VMC and SM transition densities for $A=6$, $A=10$, and $A=12$ nuclei to extract the contact values. The short-range components of the transition densities are modeled according to Eqs.~\eqref{eq:rho_F_GCF} and \eqref{eq:rho_GT_GCF} and continuously matched to re-scaled SM results, so that the long-range part $C^{0\nu}_\alpha(r)$ is  fully specified. 

While for light nuclei $A\leq12$ WSS and WSW clearly improve the transition densities in Eq.~\eqref{eq:tran_dens} in relation with the VMC results, the HO and WS radial wave functions lead to very similar results in $^{48}$Ca, see Fig.~\ref{fig:rho_F_A48_HO_WS}. Based on this observation, for $A\geq48$ nuclei our SM transitions are obtained with HO orbitals. On the other hand the SM transition densities for $A=6$, $A=10$, and $A=12$ used to extract the contact ratios against AV18+UX VMC results are always carried out with WS single-particle states. Specifically, we denote the HO results for heavy nuclei ``HO(S)'' or ``HO(W)'' depending on whether the WSS or WSW parametrization is used to extract the contacts from light-nuclei transitions.

\begin{figure}[!htbp]
\includegraphics[width=\linewidth]{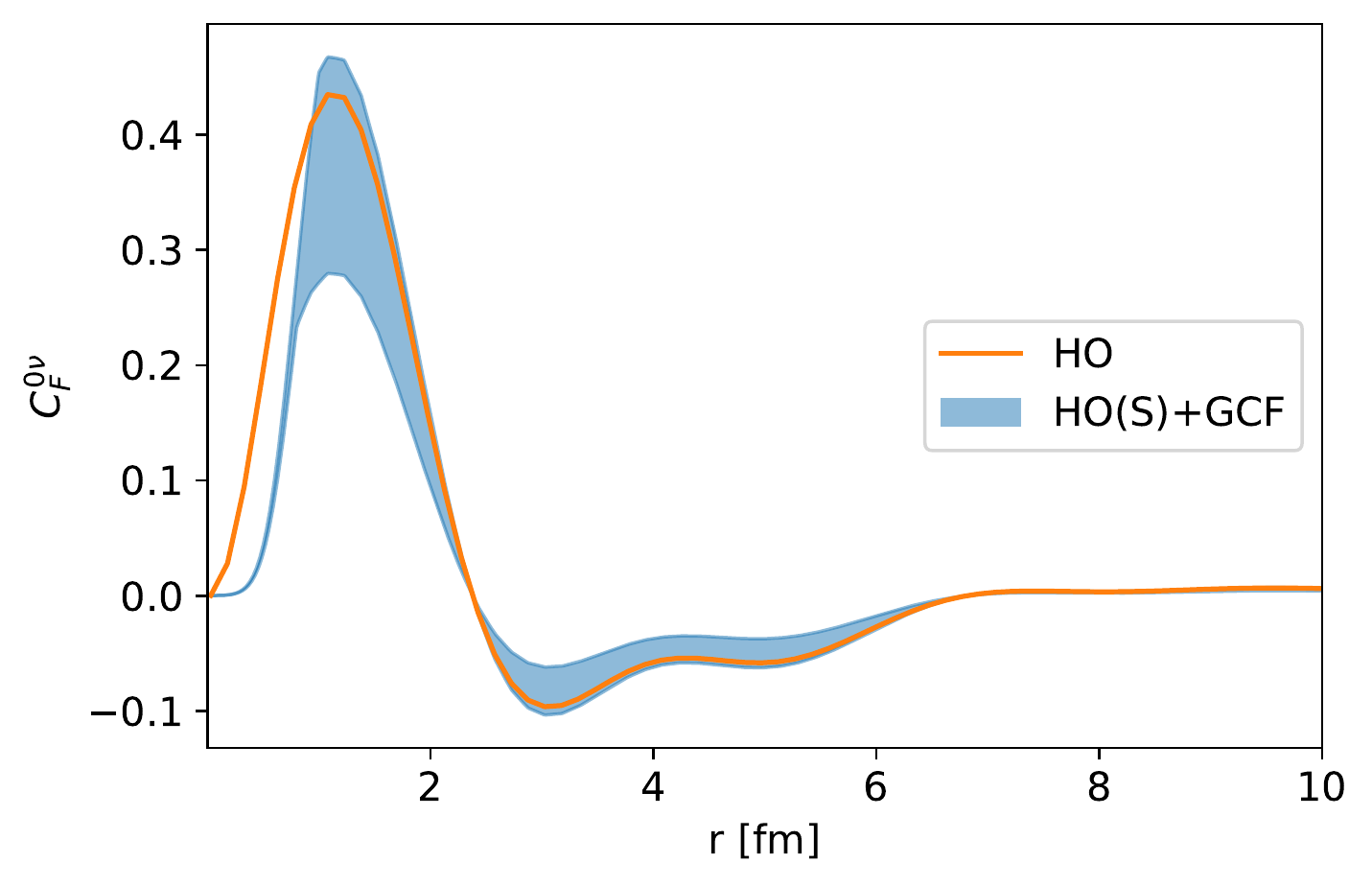}
\includegraphics[width=\linewidth]{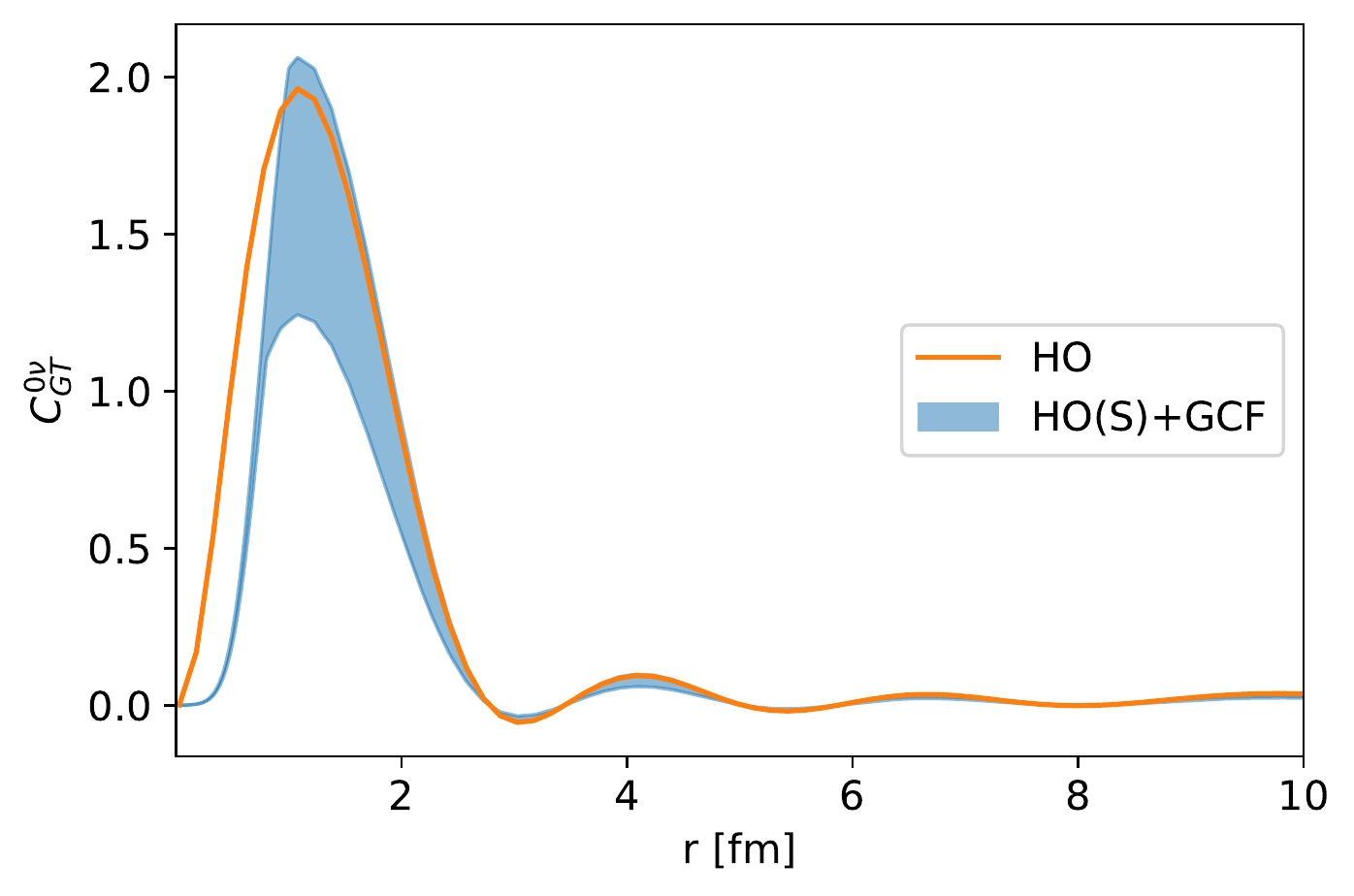}
\includegraphics[width=\linewidth]{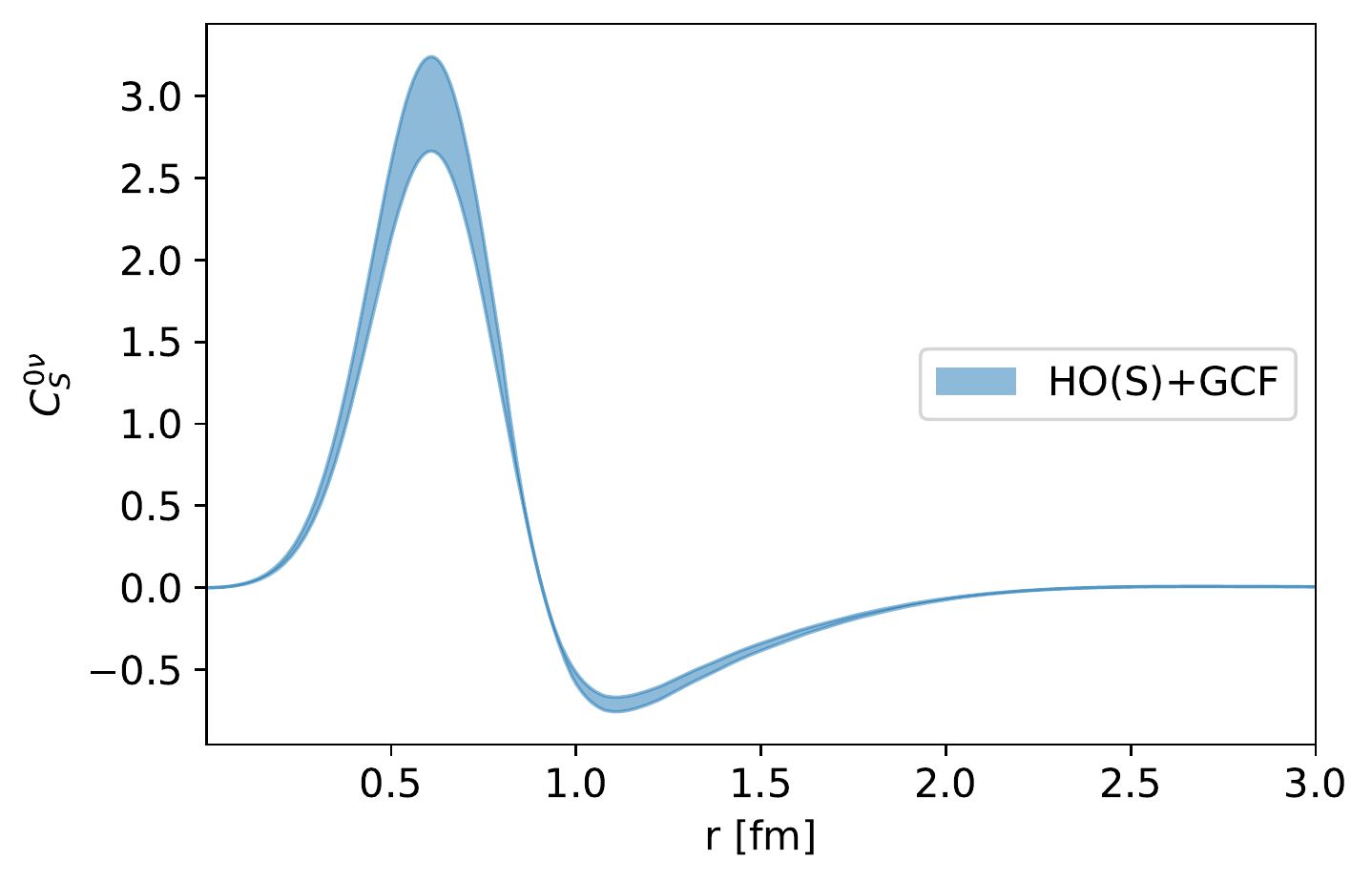}
\caption{Transition densities $C_F^{0\nu}$ (upper panel), $C_F^{0\nu}$ (middle panel), and $C_{S}^{0\nu}$ (lower panel) for the $^{76}\textrm{Ge}\rightarrow ^{76}\textrm{Se}$ decay  obtained with the SM with HO orbitals (orange line) and the GCF-SM (blue band). \label{fig:C_F_A76}}
\end{figure}

 Figure~\ref{fig:C_F_A76} illustrates the differences between SM and GCF-SM transition densities for the $^{76}\textrm{Ge}\rightarrow\, ^{76}\textrm{Se}$ decay, covering the F, GT, and SR operators. The short-range behavior of the SM is modified in a consistent fashion as in light nuclei, and reflects the underlying realistic nuclear potential. Analogously to Fig.~\ref{fig:ME_light}, we do not report the SM transition densities for the SR operator, as the corresponding NMEs are about 7 times larger than the GCF-SM values.

\begin{figure}[t]
\includegraphics[width=\linewidth]{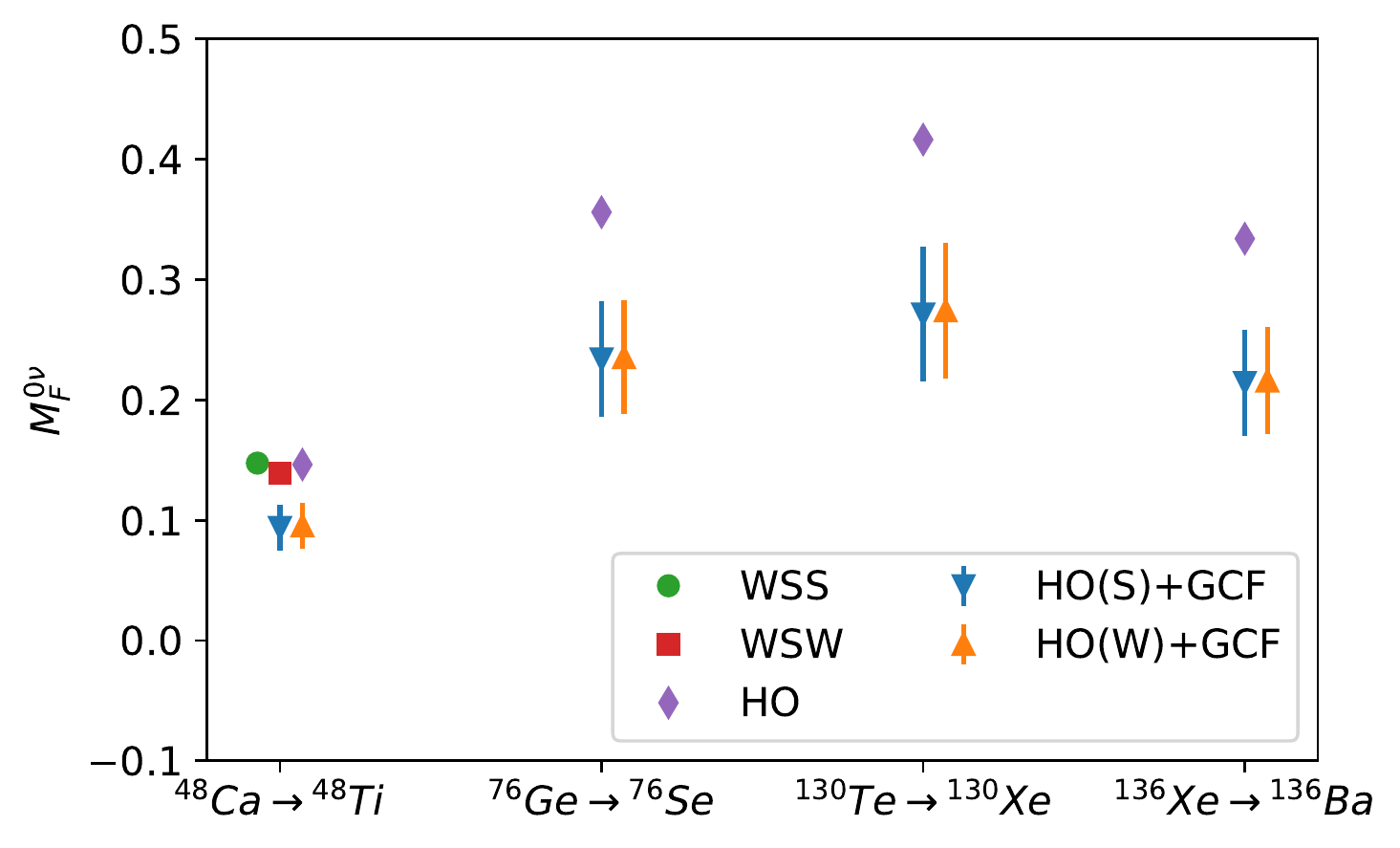}
\includegraphics[width=\linewidth]{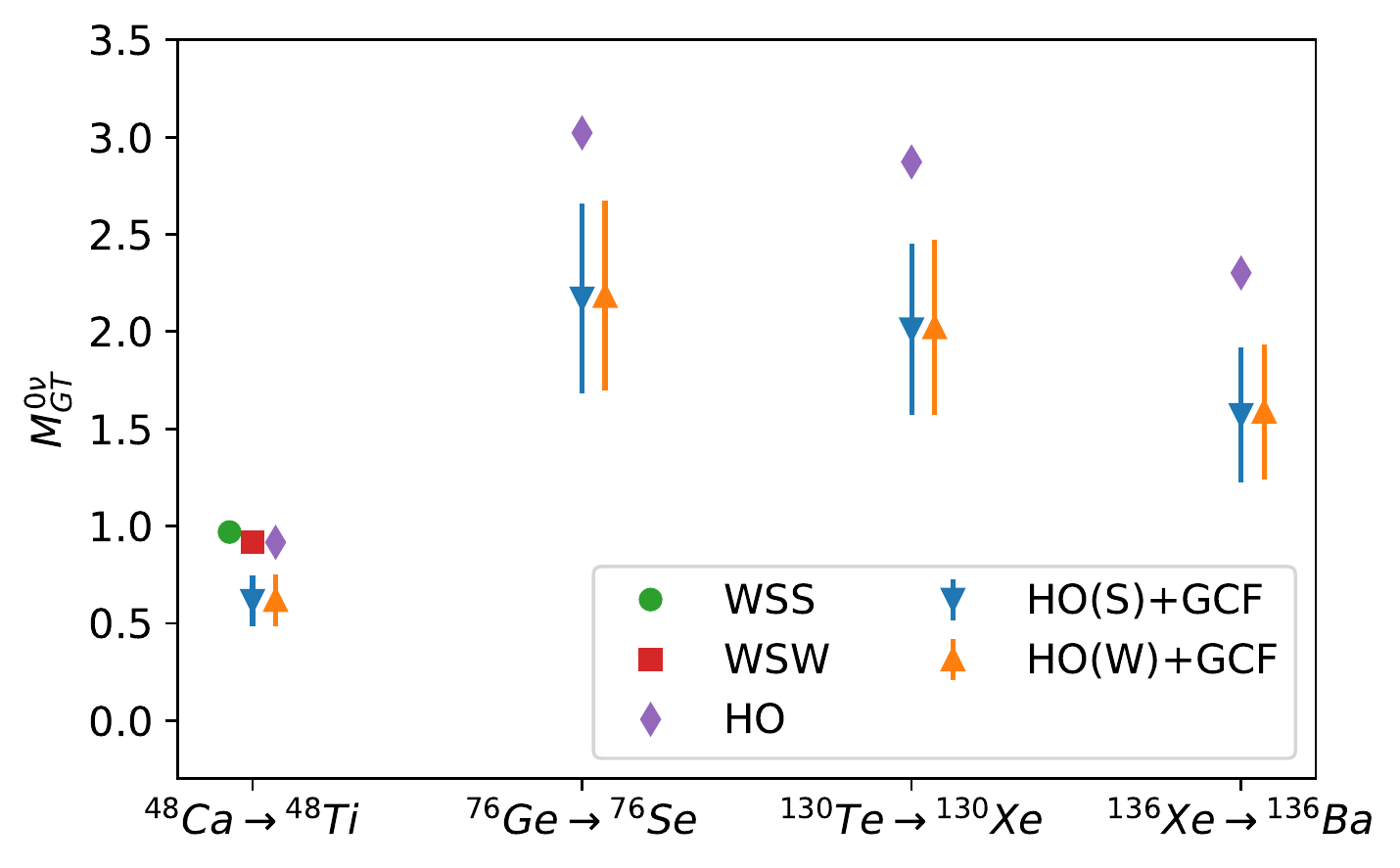}
\includegraphics[width=\linewidth]{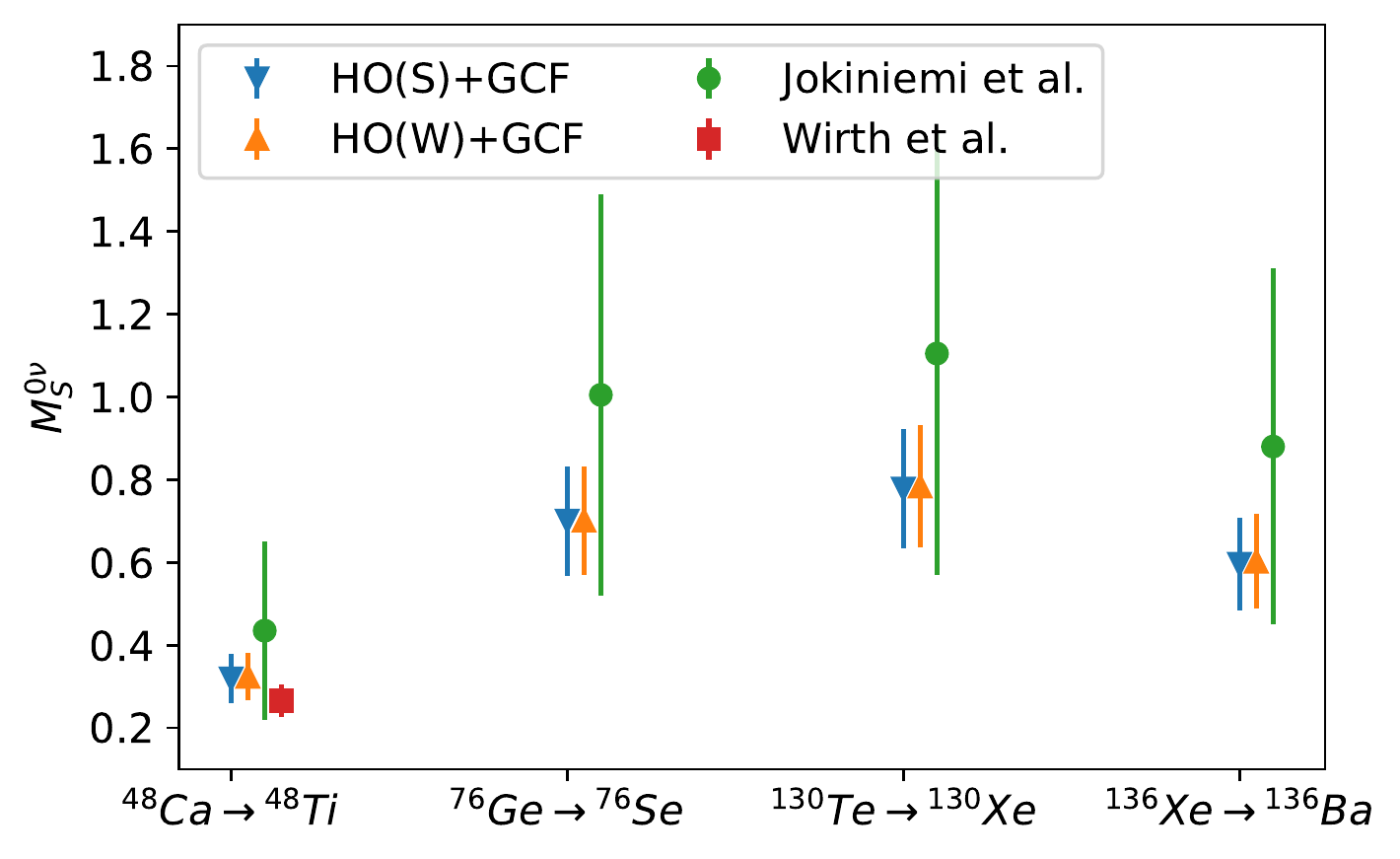}
\caption{NMEs of the F (upper panel), GT (middle panel), and SR (lower panel) operators using the GCF-SM and the SM approaches,~\label{fig:ME_F_GT_heavy} compared to the SR results of Jokiniemi {\it et al.}~\cite{JOKINIEMI2021136720} and Wirth {\it et al.}~\cite{Wirth:2021pij}.~\label{fig:ME_SR_heavy}}
\end{figure}

The GCF-SM results for the F, GT and SR matrix elements for $A=48$, $A=76$, $A=130$ and $A=136$ are displayed in Fig.~\ref{fig:ME_F_GT_heavy} and their numerical values are listed in Table~\ref{tab:MEs_heavy}. We supplement these predictions with conservative estimates for the uncertainties associated to the matching procedure and the extraction of the contact ratios as described in Sec~\ref{sec:light}. Our results indicate that the F and GT matrix elements are reduced by about $20\%-45\%$ compared to the conventional SM calculations due to the additional SRC introduced via the GCF.

The bottom panel of Fig.~\ref{fig:ME_SR_heavy} shows that the value of the short-range NME is significantly smaller in $A=48$ than in heavier nuclei, a trend which is similar when we replace the SR transition potential with the one corresponding to the NV-Ia* interaction---$M_S^{0\nu}$ only changes by about $20\%$. Figure~\ref{fig:ME_SR_heavy} also indicates that our short-range NMEs are in general smaller but consistent within error-bars with the SM results by Jokiniemi {\it et al.}~\cite{JOKINIEMI2021136720}, which cover a wider range of SR transition potentials---not including the AV18 one we use---and are further corrected by the SRC parametrization of Ref.~\cite{Simkovic2009}. In contrast, we only use one SR potential and just include uncertainties associated to the matching procedure and the extraction of the contact ratios.
The $M_S^{0\nu}$ values obtained with the QRPA by Jokiniemi {\it et al.} are somewhat larger than ours. Remarkably, our SR NME for $^{48}$Ca is in good agreement with the in-medium similarity renormalization group (IMSRG) combined with the generator coordinate method (IM-GCM) {\it ab-initio} result of Wirth {\it et al.}~\cite{Wirth:2021pij}. This is particularly interesting since they use a different nuclear interaction and also a different procedure for determining the SR coupling $g^{\text{NN}}_{\nu}$.

\begin{table}[t]
\begin{center}
\begin{tabular}{c | c | c c c c c  }
\hline
\hline
 Transition & Method & F     & GT        & SR  \\
\hline
\multirow{7}{*}{$^{48}$Ca $\rightarrow$ $^{48}$Ti} &
WSS               & 0.147    & 0.969     & 2.360   \\
&WSW              & 0.139    & 0.917     & 2.043   \\
&HO               & 0.146    & 0.916     & 2.356   \\
&WSS+GCF          & 0.09(2)  & 0.66(14)  & 0.32(6) \\
&WSW+GCF          & 0.09(2)  & 0.62(13)  & 0.27(5) \\
&HO(S)+GCF        & 0.09(2)  & 0.62(13)  & 0.32(6) \\
&HO(W)+GCF        & 0.10(2)  & 0.62(13)  & 0.32(6) \\
\hline 
\multirow{3}{*}{$^{76}$Ge $\rightarrow$ $^{76}$Se} &
HO                & 0.356    & 3.022     & 5.247  \\
&HO(S)+GCF        & 0.23(5)  & 2.17(49)  & 0.70(13) \\
&HO(W)+GCF        & 0.24(5)  & 2.18(49)  & 0.70(13) \\
\hline 
\multirow{3}{*}{$^{130}$Te $\rightarrow$ $^{130}$Xe} &
HO                & 0.416    & 2.873      & 5.804 \\
&HO(S)+GCF        & 0.27(6)  & 2.01(44)  & 0.78(14) \\
&HO(W)+GCF        & 0.27(6)  & 2.02(45)  & 0.78(15) \\
\hline 
\multirow{3}{*}{$^{136}$Xe $\rightarrow$ $^{136}$Ba} &
HO                & 0.334    & 2.302     & 4.564 \\
&HO(S)+GCF        & 0.21(4)  & 1.57(35)  & 0.60(11) \\
&HO(W)+GCF        & 0.22(4)  & 1.59(35)  & 0.60(11) \\

\end{tabular}
\caption{Fermi, Gamow-Teller, and short-range NMEs for the $^{48}$Ca $\rightarrow$ $^{48}$Ti, $^{76}$Ge $\rightarrow$ $^{76}$Se, $^{130}$Te $\rightarrow$ $^{130}$Xe and $^{136}$Xe $\rightarrow$ $^{136}$Ba transitions using the SM (HO, WSS, WSW) and the GCF-SM with different WS orbitals to fix the values of the contact ratios. 
\label{tab:MEs_heavy}}
\end{center}
\end{table}

\begin{figure}[t]
\includegraphics[width=\linewidth]{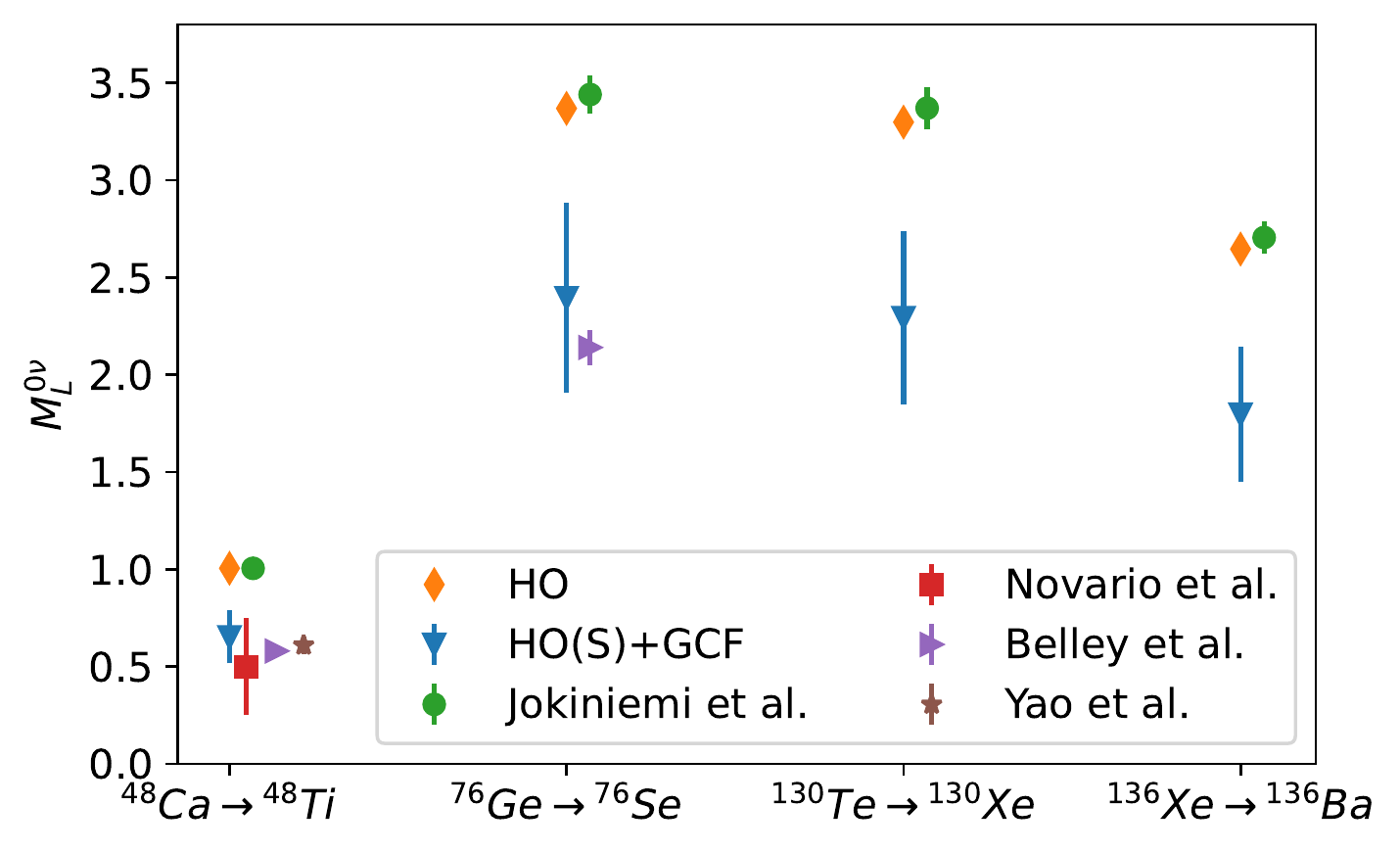}
\caption{Long-range matrix element $M_L^{0\nu}$ calculated with the combination of the GCF and the SM (blue), the SM without (orange) and with SRCs from Jokiniemi {\it et al.} (green) \cite{JOKINIEMI2021136720}, the CC from
Novario {\it et al.} (red) \cite{Novario:2020dmr}, the VS-IMSRG from Belley {\it et al.} (purple) \cite{Belley:2020ejd}, and the IMSRG+GCM from Yao {\it et al.} (brown) \cite{Yao2020}.~\label{fig:ME_L_heavy}}
\end{figure}

Eventually, the total $0\nu\beta\beta$ decay NME is the sum of the long-range term $M_L^{0\nu}=M_{GT}^{0\nu}+M_F^{0\nu}+M_T^{0\nu}$ and the short-range matrix element $M_S^{0\nu}$. As discussed in Section~\ref{sec:many_body}, we evaluate the relatively small $M_T$ contribution within the standard SM, associated to a conservative $50\%$ uncertainty. Figure \ref{fig:ME_L_heavy} presents our results for $M_L^{0\nu}$, highlighting that the GCF-SM reduces the value of $M_L^{0\nu}$ by about $15\%-40\%$ compared to the original SM calculations.
Therefore, our approach introduces a much larger SRC effect than the one from typical SRC parametrizations such as the one from Ref.~\cite{Simkovic2009} used in SM $0\nu\beta\beta$ studies, see the very small difference between the SM (HO) and Jokiniemi {\it et al.} results from Ref.~\cite{JOKINIEMI2021136720}.
Figure~\ref{fig:ME_L_heavy} also compares our NMEs with the
{\it ab-initio} results of Novario {\it et al.} \cite{Novario:2020dmr} using the Coupled Cluster (CC) method  and of Yao {\it et al.} \cite{Yao2020} using the IM-GCM approach for $^{48}$Ca and of Belley {\it et al.} \cite{Belley:2020ejd} using the valence space IMSRG (VS-IMSRG) method for $^{48}$Ca and $^{76}$Ge. Our long-range NMEs are in very good agreement with all the {\it ab initio} results for $^{48}$Ca even though these calculations use different nuclear interactions, and are also consistent with the VS-IMSRG for $^{76}$Ge. This good agreement supports our predictions for $^{130}$Te and $^{136}$Xe, for which {\it ab initio} NMEs are not available.

Comparing Figs.~\ref{fig:ME_SR_heavy} and \ref{fig:ME_L_heavy}, it is apparent that the SR term contributes significantly to the total NME: $M_S^{0\nu}$ is around $35\%-60\%$ of $M_L^{0\nu}$ for $A=48$---in good agreement with Ref.~\cite{Wirth:2021pij}---and around $25\%-40\%$ for the heavier nuclei. Since $M_S^{0\nu}$ has the same sign as $M_L^{0\nu}$, the total matrix element would be enhanced, also in agreement with Ref.~\cite{Wirth:2021pij}. Despite the differences between our results and the SM ones of Jokiniemi {\it et al.}, the relative importance of the SR term is overall similar, while the QRPA predicts somewhat larger larger ratio values~\cite{JOKINIEMI2021136720}.

It is important to highlight the differences between the GCF-SM and previous attemps to include SRCs into the SM and other approaches based on regularized interactions.
Some of the correlation functions that produce larger effect of SRCs, e.g. those by Miller and Spencer~\cite{Miller_Spencer_1976}, were criticized because they lead to violation of isospin symmetry: the integral $\int_0^\infty r^2 \rho_F(r) dr\neq0$ when the isospin of the initial and final state is different~\cite{Engel2011}. Given the relatively large effects of SRCs, one might think that the GCF-SM approach can suffer from the same shortcomings. Ref.~\cite{Engel2011} also claims that, in order to respect isospin symmetry, correlation functions should peak around $r\simeq1$ fm with a value above 1 to compensate for the reduction of probability at short distances. This behavior eventually leads to a relatively small effect on the NME values.
In contrast, to match the VMC and SM results an appropriate correlation function should be defined as the ratio of the corresponding transition densities---by construction, multiplying the SM results by this correlation function reproduces the VMC one. By comparing the VMC and SM transition densities presented in Fig.~\ref{fig:A12}, we notice that the correlation function does not peak around $r\simeq1$ fm. Furthermore, the GCF-SM approach has a significant difference in that the SM results are re-scaled to match the short-range behavior, so that the effect of the GCF does not approach unity at long distances unlike most SRC parametrizations. This re-scaling allows the GCF-SM to compensate for the short-range reduction without a peak at $r\simeq1$ fm. Some violation of the isospin orthogonality can still be found in our actual results, but this is due to subleading corrections, like three-body correlations, and possible small differences between the SM and the exact solution at long distances. Eventually, the good agreement between the GCF-SM and VMC results in light nuclei shown in Fig.~\ref{fig:A12} demonstrates the accuracy of our method. 

In addition, most of the available SRC functions assume in their derivation a simple form for the uncorrelated wave function, like a Slater determinant in Ref.~\cite{Simkovic2009}. The latter differs from the SM wave function and therefore leads to inconsistencies when combined with SM calculations.
Likewise, the correlation function based on VMC calculations introduced in Ref.~\cite{CRUZTORRES2018304} also uses a simple function for the uncorrelated part. Further, Ref.~\cite{CRUZTORRES2018304} uses proton-proton VMC densities of a given nucleus in contrast to the transition densities involving the initial and final nuclei used in the GCF-SM approach.
In short, the GCF-SM replaces the need of introducing SRC functions by directly providing the appropriate short-range structure for any given $N\!N$ interaction.

\section{Conclusions}
\label{sec:conclusions}
We have introduced a novel protocol based on the GCF that combines SM and QMC methods to compute $0\nu \beta\beta$ decay nuclear matrix elements of heavy nuclei relevant for experimental searches. The GCF is used to describe the transition densities at short-distances from QMC calculations, while re-scaled SM calculations are used to model the long-range components. A key role in our GCF-SM approach is played by the ``contact'' values, which determine the number of short-range correlated pairs participating in the transition densities. Assuming their model-independence---extensively verified in diagonal two-body densities---we extract the contact values of heavy nuclei combining VMC calculations of light nuclei with SM transitions of both light and heavy isotopes. We verify the accuracy of this procedure on VMC transition densities of $A=6$, $A=10$, and $A=12$ nuclei. The latter are also improved compared to earlier calculations by introducing a complete p-shell representation of the $^{12}$Be wave function. We supplement the GCF-SM predictions by conservative estimates of the uncertainties due to the matching procedure and extracting the contact ratios. 

We employ the GCF-SM to predict NMEs for $^{48}$Ca, $^{76}$Ge, $^{130}$Te and $^{136}$Xe. The long-range matrix elements  are appreciably reduced by $15\%-40\%$ with respect to the SM calculations. In particular, the impact of SRCs is significantly larger within the GCF-SM than when using relatively-soft functions to incorporate SRCs effects into the SM. In fact, our approach replaces altogether the need of using correlation functions, since, besides an overall normalization factor, the short-range behavior of the transition density is fully determined by the GCF. Remarkably, our results are consistent within error bars with recent {\it ab-initio} results for $^{48}$Ca and $^{76}$Ge, carried out within the CC, and VS-IMSRG, and IM-GCM methods. 
Further, we make GCF-SM predictions for the heaviest emitters used in $0\nu\beta\beta$ searches: $^{130}$Te and $^{136}$Xe, for which {\it ab-initio} results are not currently available. Given the agreement with the VMC in the light-nuclei sector and with other {\it ab-initio} approaches for $^{48}$Ca and $^{76}$Ge, we believe that the GCF-SM is a reliable complementary approach to calculate $0\nu\beta\beta$ NMEs and reduce their theoretical uncertainty.

The GCF-SM method is especially suitable for calculating the recently introduced leading order SR matrix element. Using a transition potential consistent with the AV18 interaction used to compute the transition densities at short distances, we find that the SR term enhances the total NME by $25\%-40\%$ in heavy nuclei, which is consistent with IM-GCM and SM estimations.
Nonetheless, our SR NMEs obtained based on the CIB of AV18
may need to be rescaled once the correct $g^{\text{NN}}_{\nu}$ coupling is determined---see Ref.~\cite{Cirigliano:2020dmx} for a recently proposed strategy using synthetic data.

A limitation of this work is the absence of two-body currents in the $0\nu\beta\beta$ decay, which are related to a consistent treatment of the transition operator. Two-body currents are necessary to reproduce $\beta$ decay matrix elements~\cite{Pastore:2017uwc,Gysbers:2019uyb}, but are not fully included in any $0\nu\beta\beta$ calculation yet, where their impact has only been estimated within a simple approximation~\cite{Menendez:2011qq,Engel:2014pha}.
However, Refs.~\cite{Pastore:2017uwc,Gysbers:2019uyb} indicate that two-body currents are relatively less important when using hard nuclear interactions characterized by high-momentum components, like AV18. In this sense, our NMEs the absence of two-body currents in our results may have a smaller impact than on the {\it ab initio} methods
relying on single-particle basis expansion that deal with softer nuclear potentials.

The GCF-SM $0\nu\beta\beta$ matrix elements presented in this work rely on VMC calculations carried out with the AV18+UX Hamiltonian, and the short-range GCF two-body function has consistently been computed with AV18. In future work we plan to study the NME dependence on the nuclear Hamiltonian of choice, including those derived within chiral effective field theory. For instance, the local chiral Norfolk two- and three-body potentials can be readily incorporated in the GCF-SM method once the VMC calculations are carried out. On the other hand, including non-local potentials would require calculating transition densities for light nuclei with suitable many-body methods, like the no-core shell model~\cite{Basili:2019gvn,Yao:2020olm}.

A further development of the method is to improve the treatment of the T matrix element. This is especially important in view of a recent {\it ab-initio} studies that find relatively large T contributions compared to the SM~\cite{Belley:2020ejd}. This development will require disentangling the different p-wave contributions at short distances and additional analyses of the model independence of contact ratios.

More generally, the GCF-SM approach, anchored on VMC calculations of light nuclei, can be applied to incorporating the effect of SRCs in a variety of nuclear quantities accessible by the nuclear SM. We envision computing the transition densities relevant for studying the role of correlations and two-body currents in single-$\beta$ decay rates. As noted in Ref.~\cite{King:2020wmp}, the two-body densities exhibit a universal, i.e. nucleus independent, behavior at short distance. Also, we plan on utilizing the GCF-SM method to analyze momentum distributions and spectral functions of nuclei of interest in the context of electron-scattering experiments and for the accelerator-based neutrino oscillation program.

\section{Acknowledgement}
The present research is supported by the U.S. Department of Energy, Office of Science, Office of Nuclear Physics, under contract DE-AC02-06CH11357, the NUCLEI SciDAC program (A.L., R.B.W.), by the Laboratory Directed Research and Development program of Los Alamos National Laboratory under project number 20210763PRD1 (R.W.) and by the ``Ram\'on y Cajal'' program with grant RYC-2017-22781, and grants CEX2019-000918-M and PID2020-118758GB-I00 funded by MCIN/AEI/10.13039/501100011033 and, as appropriate, by "ESF Investing in your future" (P.S. and J.M.).
The work of A.L. is also supported by the DOE Early Career Research Program award. Quantum Monte Carlo calculations were performed on the parallel computers of the Laboratory Computing Resource Center, Argonne National Laboratory, the computers of the Argonne Leadership Computing Facility via the INCITE grant ``Ab-initio nuclear structure and nuclear reactions'', and the 2020/2021 ALCC grant ``Chiral Nuclear Interactions from Nuclei to Nucleonic Matter''.

\bibliography{biblio}

\end{document}